\documentclass[superscriptaddress, twocolumn, amsmath, amssymb, aps, pra, floatfix]{revtex4-2}

\usepackage{graphicx}
\usepackage{dcolumn}
\usepackage{bm}
\usepackage{verbatim}
\usepackage{hyperref}
\usepackage{color}

\usepackage{enumitem}
\usepackage{mathtools}

\begin{document}

\title{Spin-charge correlations in finite one-dimensional multi-band Fermi systems}

\author{J. M. Becker}
\affiliation{Center for Optical Quantum Technologies, Department of Physics, University of Hamburg, 
Luruper Chaussee 149, 22761 Hamburg Germany}
\author{G. M. Koutentakis}
\affiliation{Institute of Science and Technology Austria (ISTA), Am Campus 1, 3400 Klosterneuburg, Austria}
\author{P. Schmelcher}
\affiliation{Center for Optical Quantum Technologies, Department of Physics, University of Hamburg, 
Luruper Chaussee 149, 22761 Hamburg Germany}
\affiliation{The Hamburg Centre for Ultrafast Imaging,
University of Hamburg, Luruper Chaussee 149, 22761 Hamburg, Germany}

\date{\today}

\begin{abstract}
We investigate spin-charge separation of a spin-$\frac{1}{2}$ Fermi system confined in a triple well where multiple bands are occupied. We assume that our finite fermionic system is close to fully spin polarized while being doped by a hole and an impurity fermion with opposite spin. Our setup involves ferromagnetic couplings among the particles in different bands, leading to the development of strong spin-transport correlations in an intermediate interaction regime. Interactions are then strong enough to lift the degeneracy among singlet and triplet spin configurations in the well of the spin impurity  but not strong enough to prohibit hole-induced magnetic excitations to the singlet state. Despite the strong spin-hole correlations, the system exhibits spin-charge deconfinement allowing for long-range entanglement of the spatial and spin degrees of freedom.
\end{abstract}

\maketitle

\section{Introduction}

The interplay of magnetic properties and charge carrier transport is fundamentally important for understanding the behavior of a large class of strongly correlated materials \cite{Kontani2023,Nakata2021,SingletonJames2022,Tsvelik22,Khomskii2009,Matsuda2017}.
A concrete example of this are hole-doped Mott insulators where the dressing of the hole carriers by local excitations of the gas is put forward as the origin of important phenomena such as the formation of the pseudogap phase \cite{Sacuto22,Ferrero2018}, magnetic polarons \cite{GrossBlochDemler2019,DemlerBohrdt21,Miyamoto2018} and high-temperature superconductivity \cite{Tremblay2021,Phillips2020}.
Ultracold atom simulators have been employed to unveil the complex physics of the interplay of magnetism and conductivity especially focusing on the two-dimensional case \cite{Zwierlein2019, DemlerBohrdt21, DemlerBlochGross2017, Bohrdt2022, GrossBloch2021, GrossBlochDemler2019, DemlerGreiner2021, BohrtHomeier2021, Hirthe2023}.

The interplay of magnetism and transport in the one-dimensional case is thought to be much simpler, due to the effect of spin-charge separation \cite{Crommie2022,Tserkovnyak07,Piroli2021,MurakamiTakayoshi2023, OgataShiba1990}.
Spin-charge separation implies that the spin and particle excitations spatially propagate with different velocities while not interacting with one another.
This effect was experimentally identified in lattice setups \cite{GrusdtBlochGross2020} and it is considered to be a generic property of Luttinger liquids, being the low-energy description of one-dimensional interacting Fermi systems \cite{Hulet2022}. As a result of the general assumption of spin-charge separation, the possibility of a more involved interplay among magnetism and transport in one dimension is hardly discussed in the literature.

Recently, important advances in the understanding of magnetic properties in one-dimensional setups have been made in both the weak \cite{KollerMundinger2015, KollerWall2016, Koutentakis2019, KoutentakisMistakidis2020} and strong \cite{VolosnievFedorov2014, DeuretzbacherBecker2014, CuiHo2014, YangGuan2015, LevinsenMassignan2015, Landman2016, Jochim2015} interaction regimes. This is enabled by the experimental availability of controllably preparing few-body fermionic ensembles, allowing for the detailed study of the underlying microscopic mechanisms~\cite{SerwaneZuern2011, Jochim2015, WenzZuern2013, MurmannBergschneider2015, BergschneiderKlinkhamer2018, BayhaHolten2020}.
Such studies are aided by the available powerful theoretical tools referring to \textit{ab initio} approaches \cite{SowinskiLewenstein2013, BrandtYannouleas2015, CaoBolsinger2017} and effective spin-chain models \cite{KollerMundinger2015, KollerWall2016, Koutentakis2019, KoutentakisMistakidis2020, VolosnievFedorov2014, DeuretzbacherBecker2014, CuiHo2014, YangGuan2015, LevinsenMassignan2015, Landman2016}. 
Furthermore, theoretical studies that attempt to connect atom transport properties to magnetic phenomena have been performed \cite{VolsnievPetrosyan2015, MarchukovVolosniev2016, BarfknechtFoerster2019}. However, within these works spin-charge separation is generally employed as an {\it a priori} assumption.
 \begin{figure}
    \centering
    \includegraphics[width=1.0\columnwidth]{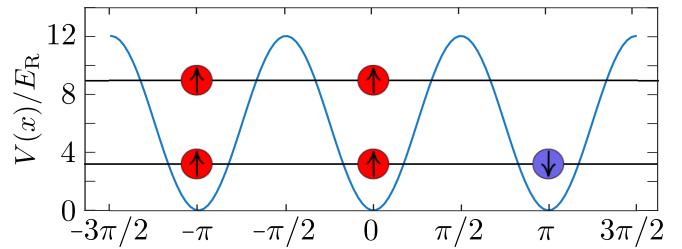}
    \caption{Initial state of the system consisting of four spin-$\uparrow$ particles (red) and one spin-$\downarrow$ particle (blue) trapped in an one-dimensional triple-well potential $V(x)$ (see text).}
    \label{schematic}
\end{figure}

Recent studies of ferromagnetic phenomena in one-dimensional setups motivate a more involved relation between spin and particle transport~\cite{Koutentakis2019, KoutentakisMistakidis2020}. In particular, these works unveiled the interplay of kinetic energy promoted anti-ferromagnetism and interaction-driven ferromagnetic order in the excited states of fermionic ensembles~\cite{KoutentakisPhD,KoutentakisMistakidis2020}. This is captured by models, where apart from the spatial coordinate, the energy band of the different involved states contributes as an additional degree of freedom \cite{KoutentakisMistakidis2020}. As a consequence, the possibility of coupling spin and charge excitations cannot be certainly ruled out, allowing for transport beyond the paradigm of spin-charge separation. However, whether this prospect can be realized in a one-dimensional system or if it is an artifact of this effective description is yet to be explored.

In this paper, we demonstrate correlated spin and particle transport within a one-dimensional setup by employing a spin-$\frac{1}{2}$ fermionic system confined in a triple-well potential. In particular, we consider a polarized ensemble of spin-$\uparrow$ fermions in a double well subsystem and a single spin-$\downarrow$ impurity fermion in a third well (see Fig.~\ref{schematic}). Notice, that this system is doped by a single hole (initially located in the third well) allowing us to probe the interplay of spin and charge. Within an appropriate interaction strength interval, the dynamics of the system indicates that the tunneling of the hole couples to distinct spin configurations. Thus, it interacts with the magnetic order emanating in the triple well, indicating the development of spin-charge correlations. The correlated dynamics of this spinor-fermion system is described on an {\it ab initio} level by employing the multilayer multiconfiguration time-dependent Hartree method for atomic mixtures (ML-MCTDHX)~\cite{CaoBolsinger2017} and the involved mechanisms are elucidated in terms of a phenomenological model. The $tJU$ spin-chain model~\cite{KoutentakisMistakidis2020} provides the bridge between these two paradigms.

Beyond the interaction strength interval, where spin-charge correlations arise, for stronger or weaker repulsion, spin-charge separation is re-established. In case of very weak interactions, this is manifested by uncorrelated tunneling of the spin-$\uparrow$ particles on the second band, since the ferromagnetic exchange is negligible.  On the other hand, for very strong interactions, correlations among the hole and the spin-$\downarrow$ fermion persist, but the latter behaves as a nonmagnetic impurity and the system exhibits ferromagnetic correlations for all times. We note that the two (quasi)particles cannot bind due to the repulsive character of the hole-spin-$\downarrow$ interactions, which results in the development of long-ranged correlations among them. This effect is observable when larger lattice setups are considered and could be exploited to imprint spin-transport correlations in spintronic devices~\cite{vanderStraten2013}.

This paper is structured as follows. In Sec.~\ref{sec:systemdescription}, we describe our setup and discuss its spin symmetries. Subsequently, in Sec.~\ref{sec:absentSC} we present the many-body dynamics and show that spin-charge separation does not hold in this multi-band system. We introduce the spin coupling mechanism that can explain this effect and demonstrate its basic dynamical implications within a phenomenological model. An effective $tJU$ model is introduced in Sec.~\ref{sec:effectivemodel}, which is compared to the {\it ab initio} results and used to expose the spin-state coupling mechanism. Finally, we conclude in Sec.~\ref{sec:conclusion} and provide an outlook on possible future perspectives. Our numerical method is illustrated in Appendix~\ref{sec:mlx}. Appendix~\ref{sec:non-interacting-effective} provides the derivation of the effective model in case of non-interacting particles while Appendix~\ref{app:effective} provides the effective interaction Hamiltonian within the $tJU$ model. Appendix~\ref{app:doublons} demonstrates the effect of doublons in the dynamics.

\section{Setup, Hamiltonian, and State Preparation}
\label{sec:systemdescription}
We consider a spin-$\frac{1}{2}$ system consisting of $N_\uparrow=4$ spin-$\uparrow$ fermions interacting with a single $N_\downarrow=1$ spin-$\downarrow$ fermion confined in a one-dimensional triple-well potential $V(x)=V_0\, \mathrm{sin}^2(kx)$. 
The latter can be realized in an experiment by an optical lattice with wavelength $\lambda=2\pi/k$. In order to restrict the system to three wells, we impose hard wall boundary conditions at $x=\pm 3\pi/(2 k)$. The inverse lattice wave number $k^{-1}$ and the recoil energy $E_\mathrm{R}=\hbar^2 k^2/(2 m)$ constitute the characteristic length and energy scales, while the corresponding timescale reads as $t_{\rm R} = \hbar/E_{\rm R}$. In what follows, all mentioned numerical values are to be understood in the units derived from $k^{-1}$, $E_{\rm R}$ and $t_{\rm R}$. Subsequently, we choose a potential height of $V_0=12 E_{\rm R}$ realizing deep wells that allow particles to be well-localized. Finally, let us note that the two different spin-states can be experimentally implemented by using ultracold $^6$Li-atoms in the two lowest hyperfine states $|F=1/2, m_F=-1/2\rangle$ and $|F=1/2, m_F=+1/2\rangle$ \cite{GrossBlochDemler2019}.

\subsection{Many-body Hamiltonian}
Our system is described by a many-body Hamiltonian that can be separated into a non-interacting part for both species $\sigma\in\{\uparrow,\downarrow\}$ and a part describing the interactions between them: $\hat{H}= \sum_{\sigma\in\{\uparrow,\downarrow\}}\hat{H}_\sigma+\hat{H}_I$. The underlying single-species Hamiltonian reads as
\begin{equation}
    \hat{H}_\sigma = \int \mathrm{d}x~ \hat{\psi}^\dagger_{\sigma}(x) \left(- \frac{\hbar^2}{2m}\frac{\mathrm{d^2}}{\mathrm{d}x^2}+ V(x) \right)\hat{\psi}_{\sigma}(x),
    \label{Hamilton_mlx_bare}
\end{equation}
where $\hat{\psi}_{\sigma}(x)$ is the fermionic field operator annihilating a spin-$\sigma$ fermion at position $x$.
The interactions are governed by
\begin{equation}
\begin{split}
        \hat{H}_I &= \frac{g}{2} \int \mathrm{d}x~ :\hat{n}^2(x): \\
        &= g \int \mathrm{d}x~ \hat{\psi}^\dagger_{\uparrow}(x)\hat{\psi}^\dagger_{\downarrow}(x)\hat{\psi}_{\downarrow}(x)\hat{\psi}_{\uparrow}(x),
    \label{Hamilton_mlx_int}
\end{split}
\end{equation}
where $\hat{n}(x) = \hat{\psi}^\dagger_{\uparrow}(x)\hat{\psi}_{\uparrow}(x) + \hat{\psi}^\dagger_{\downarrow}(x)\hat{\psi}_{\downarrow}(x)$ is the spin-independent density operator and $:\hat{O}:$ implies normal ordering of the operator $\hat{O}$.
Since we consider ensembles of non-dipolar ultracold atoms, their interactions are captured by effective one-dimenstional (1D) $s$-wave contact interactions. The interaction strength $g$ is linked to the (3D) $s$-wave scattering length and the  transverse confinement length. Thus, it is experimentally tunable via Fano-Feshbach and confinement-induced resonances \cite{Olshanii98,JulienneTiesinga,BergemannOlshanii}.

The total Hamiltonian $\hat{H}$ commutes with the spin operators $\hat{S}_{\pm}=\hat{S}_x\pm i \hat{S}_y$ and $\hat{S}_z$, since both $\hat{H}_{\uparrow} + \hat{H}_{\downarrow}$ and $\hat{H}_I$ are spin-independent operators. Recall that for an itinerant system the individual spin operators are defined as $\hat{S}_i=\frac{1}{2}\int \mathrm{d}x \sum_{\sigma,\sigma^\prime} \hat{\psi}^\dagger_\sigma(x) \pmb{\sigma}^i_{\sigma,\sigma^\prime}\hat{\psi}_{\sigma^\prime}(x)$, where $\pmb{\sigma}^i$ are the Pauli matrices with $i \in \{x,y,z\}$. Furthermore, $\hat{H}$ is invariant under rotations in spin-space, associated with its SU(2)-symmetry stemming from its commutation with the total spin operator $\hat{S}^2=\hat{S}_{+} \hat{S}_{-}+ \hat{S}_z(\hat{S}_z -1)$. This implies that the total spin $S$ of the system is conserved throughout the dynamics. 

We analyze the time-evolution of the system for a repulsive interaction regime of $g\in \left[0,5\right]$. In the non-interacting case $g=0$, the particles are expected to tunnel among the wells similarly to isolated particles under the influence of the $V(x)$ potential. The corresponding tunneling frequencies are related to the eigenenergy differences within the quasi-degenerate triplet of eigenstates forming the precursor of Bloch bands of the translational invariant system (see Appendix~\ref{sec:non-interacting-effective}). In contrast, non-vanishing interactions $g \neq 0$ introduce a coupling between the spin-$\uparrow$ and spin-$\downarrow$ particles, possibly resulting in the development of correlations that qualitatively alter the dynamics of the system.

To track the system dynamics, we employ the variational \textit{ab initio} method ML-MCTDHX, which allows us to take all relevant correlations into account (see also Appendix~\ref{sec:mlx}).

\subsection{Initial state}
Our setup involves the two lowest energy bands, $b=0,1$, of the potential $V(x)$. This inclusion provides additional degrees of freedom when compared to one-dimensional (lowest band) Hubbard models where spin-charge separation is present \cite{GrusdtBlochGross2020, MurakamiTakayoshi2023, OgataShiba1990}. Typically, in order to describe the lattice sites on those bands in translationally invariant discrete systems, e.g. in crystals, one makes use of Wannier states, which form a well-localized basis set. In our case, the system has a limited spatial extent and thus the discrete translational invariance is broken. However, due to the large depth of the potential wells, here $V_0 = 12 E_{\rm R}$, we can define the associated Wannier states $\phi_s^b(x)$ despite the hard-wall boundaries on the potential, that are localized in either the left (L), middle (M) or right (R) well $s \in \{\mathrm{L},\mathrm{M},\mathrm{R}\}$ (for details see Appendix~\ref{sec:non-interacting-effective}). These states are used to express the initial state of our system (see Fig.~\ref{schematic}) reading as
\begin{equation}
\begin{split}
        |\Psi(0)\rangle = \left(\prod_{b=0,1} \prod_{s\in \{\mathrm{L}, \mathrm{M}\}} \int  \mathrm{d}x\phi_s^b(x) \hat{\psi}^\dagger_\uparrow(x)\right) \\ \times \left( \int  \mathrm{d}x \phi_\mathrm{R}^0(x)\hat{\psi}^\dagger_\downarrow(x)\right) |0\rangle .
\end{split}
\label{initialstate_equ}
\end{equation}
The unoccupied Wannier state in the upper energy band of the right well, $b = 1$, will be referred to as the ``hole''. Experimentally, such states can be generated by preparing polarized gases in different spatial locations and then loading them in the triple-well setup. Since each particle within a spin-polarized Fermi gas behaves independently of the others, any transfer approach that works for a single particle (see e.g. \cite{KoepsellHirthe2020}) can be used for a polarized gas. In this case, loading the second band of the lattice is not challenging since two particles with the same spin cannot occupy the same state, so two spin-$\uparrow$ particles loaded in the same well would have to occupy energetically distinct bands. When the energy deposited into the system by the transfer process is low, those should predominantly be $b=0$ and $b=1$, realizing Eq.~\eqref{initialstate_equ}. Magnetic field gradients, detuning the spin-$\uparrow$ from the spin-$\downarrow$ states, can be employed to inhibit spin dynamics of the ensemble when the particles of unlike spin are brought together after the transfer is completed. To initiate the system dynamics, these gradients are lifted such that the single-particle states of different spin become degenerate (see also \cite{KoutentakisMistakidis2020}).

\section{Absence of Spin-Charge Separation}
\label{sec:absentSC}
In this section, we analyze the time evolution of the system after initiating the dynamics with the state of Eq.~\eqref{initialstate_equ}. We probe spin-charge separation and observe that it does not hold for all interaction strengths. In order to explain this effect, a coupling mechanism among the spin and spatial states is proposed, the implications of which are elucidated within a phenomenological model.

\subsection{Hole dynamics}
\label{sec:holedyn_mlx}

In the case of spin-charge separation, we expect the hole, being related to charge transport, to manifest dynamics that is independent of the interaction strength, $g$, influencing spin transport. In order to probe this, we pinpoint the position of the hole in the lattice by tracking the experimentally relevant observables, 
\begin{equation}
   h_s(t) =2 - n_{s}(t),
   \label{hole_operator_MLX}
\end{equation}
where $n_s(t)$ is the occupation of the well $s \in \{\mathrm{L}, \mathrm{M}, \mathrm{R}\}$ during time evolution. These observables determine whether there are less than two particles in each well.
The number of particles in each well is determined numerically via the integral of the corresponding spin-independent one-body density,
\begin{equation}
    n_s(t) = \sum_{\sigma \in \{\uparrow, \downarrow\}}\int_{x_{i,s}}^{x_{f,s}} \mathrm{d}x \hat{\rho}^{(1)}_\sigma(x;t),
    \label{definition_ns}
\end{equation}
where the interval limits $\left[x_{i,s},x_{f,s}\right]$ corresponding to the spatial extent of the three wells, $s \in \{\mathrm{L},\mathrm{M},\mathrm{R}\}$. Namely, $x_{i, L} = -3\pi \lambda/2$, $x_{f, L} = -\pi \lambda/2$, $x_{i,M} = -\pi \lambda/2$, $x_{f,M} =\pi \lambda/2$ and $x_{i,R} = \pi \lambda/2$, $x_{f,R} = 3\pi \lambda/2$. Notice that $h_s(t)$ is directly observable in experimental setups that employ quantum gas microscopy~\cite{CheukNichols2015, HallerHudson2015, ParsonsHuber2015, GrossBakr2021}, as it does not require the experimentally challenging resolution of the occupation of different bands.

In the following, we interpret non-zero positive values of $h_s(t)$ as the probability of a hole lying on site $s$. This holds as long as doublons, where two particles of opposite spin occupy a single site on a band, do not form. The validity of this assumption is discussed in detail in Appendix~\ref{app:doublons}.

\begin{figure}
    \centering
    \includegraphics[width=1.0\columnwidth]{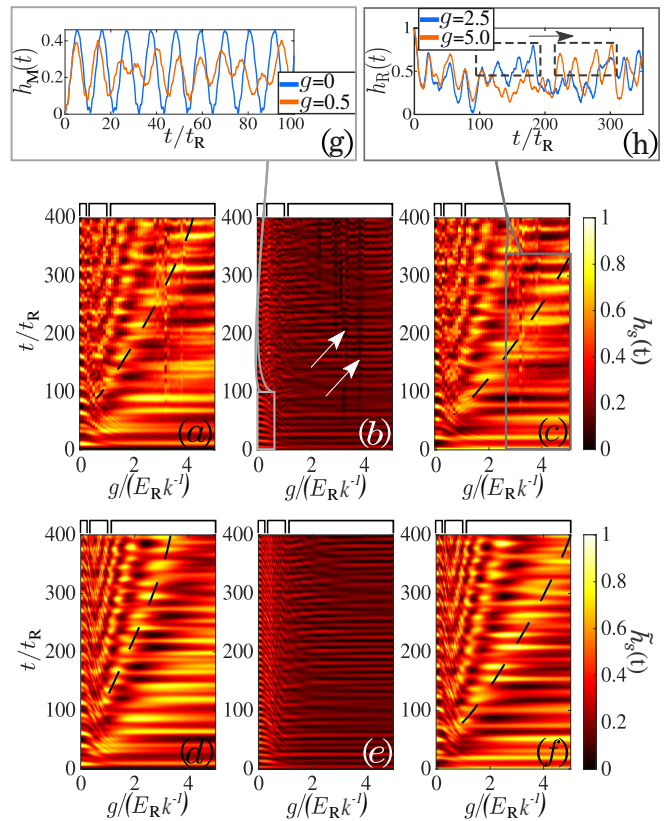}
   \caption{ Time evolution of the number of holes inside the left (a), (d), middle (b), (e) and right wells (c), (f) for the setup of Fig.~\ref{schematic} for $V_0 = 12 E_{\rm R}$. The presented results were obtained (a)--(c) by ML-MCTDHX via $h_s(t)$, see Eq.~\eqref{hole_operator_MLX} and (d)--(f) the effective $tJU$ model by employing $\tilde{h}_s(t)$ [see Eq.~\eqref{hole_operator_tJU}]. (a)--(f) Brackets above all subfigures indicate the three identified interaction regimes (see text). (g), (h) Insets compare the dynamics of the hole $h_{\rm M}(t)$ and $h_{\rm R}(t)$ for different interaction strengths. (b) White arrows  point to two exemplary resonances [compare (b) to the missing vertical stripe features in (e)]. (c) Black dashed boxes in the inset indicate the time instances when the tunneling of the hole is affected by repulsion between the hole and the spin-$\downarrow$ fermion. (a), (c), (d), (f) Dashed black lines indicate the time scales of the diffusion of the spin-$\downarrow$ fermion due to ASEI (see text).}
    \label{holecomparison}
\end{figure}

The interaction-dependent time evolution of the three observables $h_s(t)$ within the {\it ab initio} approach is depicted in Figs.~\ref{holecomparison}(a)--\ref{holecomparison}(c). Due to the given initial state (see Fig.~\ref{schematic} and Eq.~\ref{initialstate_equ}), the probability to find the hole in the right well at $t=0$ is equal to one and there are no holes in the other wells. With progressing time, the hole tunnels among the wells, primarily the left [Fig.~\ref{holecomparison}(a)] and right well [Fig.~\ref{holecomparison}(c)]. The middle well [Fig.~\ref{holecomparison}(b)] merely mediates the transfer among the other two wells and the probability to contain the hole is not pronounced. We attribute this to the broken translational symmetry of the system resulting in an energetical offset between the central and the outer wells. When the particle is in an outer well, it can tunnel only in one direction, thus its kinetic energy is larger than when it resides in the central well.

Let us now focus on the presence of spin-charge separation. Figures \ref{holecomparison}(a)--~\ref{holecomparison}(c) clearly demonstrate that increasing $g$ leads to a modification of the hole dynamics compared to $g=0$. Based on this figure, we can identify three distinct dynamical regimes (see brackets above all subfigures in Fig.~\ref{holecomparison}), which manifest in all three wells. For weak interactions, $g\leq 0.3$, the behavior of the hole is close to the non-interacting case, which is characterized by uncorrelated tunneling. Since there is no observable dependence of the hole tunneling on interactions in this regime, we conclude that spin-charge separation holds in this case. Increasing the interactions to $0.3<g<1.0$ leads to an abrupt change of the tunneling behavior of the hole. Specifically, an increase of the tunneling frequency is visible in Fig.~\ref{holecomparison}(b)  [see also Fig.~\ref{holecomparison}(g) showing $h_{\rm M}(t)$ for $g=0$ and $g=0.5$]. Note that this change is noticeable in all three wells. This alteration of the hole dynamics with $g$ implies that spin-charge separation is absent in this regime. Further increasing interactions gives rise to the strong interaction regime $g \geq 1.0$, again indicated by an abrupt change of the tunneling process, which becomes slower [see, for instance, Fig.~\ref{holecomparison}(b) for $g = 5$ compared to $g=0.5$]. In this regime, the hole dynamics regains independence of spin interactions. This is particularly visible for the middle well [see Fig.~\ref{holecomparison}(b)] since the other wells show interaction dependence for large time scales $t>100$ even in the case of $g>1$ [see Fig.~\ref{holecomparison}(a), (c), and (h)]. The reason for this $g$ dependence is related to hole and spin-$\downarrow$ interactions and will be analyzed in the following section. Finally, we can observe individual weak resonances in the dynamics inside all three wells for interactions of $g \simeq 3.0$ and $g \simeq 3.8$ [see, for example, the white arrows pointing to the vertical resonances in Fig.~\ref{holecomparison}(b) associatesd with the depletion of $h_M(t)$ during the dynamics], which we will discuss below (see Sec.~\ref{sec:effectivemodel}). While they affect $h_s(t)$ in a narrow $g$ range within the strong interaction regime, they do not alter the global dynamical behavior of the hole.

In conclusion, our results [Figs.~\ref{holecomparison}(a)--~\ref{holecomparison}(c)] indicate that the presence of spin-charge separation in our system strongly depends on the interaction strength between both spin species and cannot be \textit{a priori} assumed.

\subsection{Spin-$\downarrow$ particle dynamics}
\label{sec:spin-down-dynamics}
\begin{figure}
    \centering
    \includegraphics[width=1.0\columnwidth]{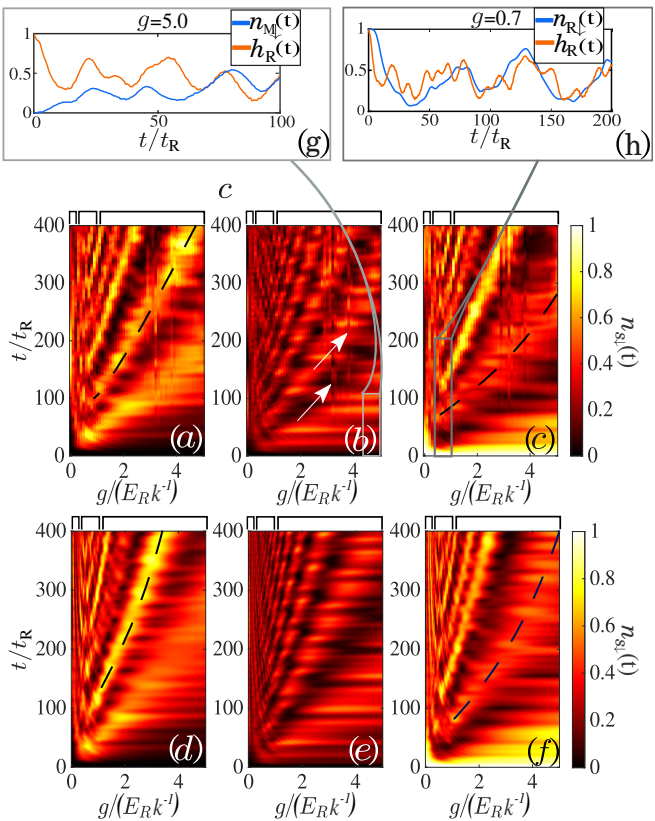}
    \caption{Time evolution of the number of spin-$\downarrow$ particles, $n_{s \downarrow}$, inside the left (a), (d), middle (b), (e) and right well (c), (f). The presented results were obtained within (a)--(c) ML-MCTDHX and (d)--(f) the effective $tJU$ model. (a)--(f) Brackets above all subfigures indicate the three identified interaction regimes (see text). (g), (h) Insets compare the dynamics of the hole $h_{\rm R}(t)$ and the spin-$\downarrow$ impurity $n_{\rm M/R \downarrow}(t)$ for $g=0.7$ [correlations between both (quasi)particles inside the same well] and for $g=5.0$ [correlations between two different wells.](b) White arrows point to two exemplary resonances [compare Fig.~\ref{spindowncomparison}(b) to Fig.~\ref{holecomparison}(b) and to the missing vertical stripe features in Figs.~\ref{holecomparison}(e) and~\ref{spindowncomparison}(e)]. (a), (c), (d), (f) Dashed black lines indicate the time scales of the diffusion of the spin-$\downarrow$ fermion due to ASEI (see text). }
    \label{spindowncomparison}
\end{figure}

In order to get a first impression of the possible dynamical correlations between the hole and spin-$\downarrow$ particle we examine the dynamics of the latter and subsequently compare to the results of $h_s(t)$. 
To this end, we depict the particle-number dynamics of the spin-$\downarrow$ particle in Fig.~\ref{spindowncomparison}. The corresponding spin-resolved particle number expectation value inside the wells $s \in \{{\rm L}, {\rm M}, {\rm R}\}$ reads 
\begin{equation}
    n_{s \downarrow}(t) = \int_{x_{i,s}}^{x_{f,s}} {\rm d}x~\rho^{(1)}_{\downarrow}(x),
    \label{spin_down_occ}
\end{equation}
with $[x_{i,s}, x_{f,s}]$ defined as for Eq.~\ref{definition_ns}.
Since the initial state is given as in Eq.~\eqref{initialstate_equ}, Figs.~\ref{spindowncomparison}(a)--~\ref{spindowncomparison}(c) indicate a probability of $n_{s \downarrow}(t = 0) = \delta_{s, {\rm R}}$.
For $t>0$, the spin-$\downarrow$ particle exhibits tunneling dynamics among the wells similarly to the hole quasiparticle [compare to Figs.~\ref{holecomparison}(a)--~\ref{holecomparison}(c)]. In contrast to the latter, the spin-$\downarrow$ dynamics show a prominent interaction dependence for all considered values of $g$. More specifically, for small interactions, $g < 0.3$, the frequency of the spin-$\downarrow$ tunneling substantially increases from the non-interacting case. This can be seen for instance in Fig.~\ref{spindowncomparison}(c) for $t<80$ and $g\leq 0.3$. The time interval for which the particle is still localized inside the right well decreases with increasing $g$ when compared to $g=0$. This effect can be attributed to spin-exchange processes \cite{KollerWall2016, Koutentakis2019, KoutentakisMistakidis2020} that allow the spin-$\downarrow$ to transfer to the excited band of the triple-well. Within $b=1$, it possesses a much larger tunneling frequency than in the ground band owing to its higher energy. For larger interactions, $g \geq 1.0$, we see that the spin-$\downarrow$ tunneling dynamics slows down as $g$ increases, visible as increased revival time scales [see Fig.~\ref{spindowncomparison}(a) for $g=2$ and $g=5$]. This feature can be understood in the following manner. As we claimed above and explicated as in Appendix D, doublons do not form in either of the bands, thus the main way that the spin-$\downarrow$ particle is transported across the triple-well is the Anderson superexchange interaction (ASEI)~\cite{Anderson1959}, via which adjacent spin-$\uparrow$ and spin-$\downarrow$ particles are exchanged via a virtual transition to a doublon state. Since the transport rate within this mechanism is inversely proportional to the interaction strength, the tunneling dynamics induced by this mechanism slows down as $g$ increases, explaining the behavior of Figs.~\ref{spindowncomparison}(a)--~\ref{spindowncomparison}(c). See also the black dashed lines in Figs.~\ref{spindowncomparison}(a) and ~\ref{spindowncomparison}(c) that illustrate the corresponding time scales of the ASEI dominated tunneling. Let us note that the dynamics analyzed above is analogous to the double-well case of Ref. \cite{Koutentakis2019}. Note that the spin-$\downarrow$ fermion shows weak resonances for $g\simeq 3.0$ and $g\simeq3.8$, just like the hole [see the white arrows pointing to vertical resonances in Fig.~\ref{spindowncomparison}(b)].

The joined analysis of Figd.~\ref{holecomparison}(a)--~\ref{holecomparison}(c) and Figs.~\ref{spindowncomparison}(a)--~\ref{spindowncomparison}(c) allows us to identify correlations between the dynamics of the (quasi)particles hinting towards the presence of a related interaction mechanism. Indeed, additional structures to the above mentioned can be observed in the transport properties of the spin-$\downarrow$ particle that can be attributed to its interaction with the hole. Example cases of this are more easily identified for strong interactions, $g > 2$, where we can observe that in addition to the ASEI dominated spin-$\downarrow$ dynamics a faster time scale associated with the hole tunneling [Fig.~\ref{holecomparison}(a) and \ref{holecomparison}(c)] is imprinted in the dynamics of spin-$\downarrow$ particle [Figs.~\ref{spindowncomparison}(a)--~\ref{spindowncomparison}(c)]. A particular example of this is the abrupt transport of the spin-$\downarrow$ particle appearing from the right [Fig.~\ref{spindowncomparison}(c)] to the middle [Fig.~\ref{spindowncomparison}(b)] well as soon as the hole density revives on the right well. This can be seen also in Fig.~\ref{spindowncomparison}(g), which shows the rough coincidence of the increasing and decreasing tendencies of $h_R(t)$ and $n_M(t)$ for $t > 10$. By inspecting the dynamics of Figs.~\ref{holecomparison}(c) and~\ref{spindowncomparison}(c), we can identify several similar events reminiscent of hole-spin-$\downarrow$ scattering processes. A similar behavior can be identified for smaller $g$, substantially modifying the hole dynamics. In particular, by comparing Figs.~\ref{holecomparison}(c) and \ref{spindowncomparison}(c) for $0.3 < g < 1$ we can see that the presence of the spin-$\downarrow$ particle in the right well is clearly correlated with an increased probability of the hole also residing in the same site [see Fig.~\ref{holecomparison}(h)]. The same is also the case for the left well [Fig.\ref{holecomparison}(a) and \ref{spindowncomparison}(a)]. 

Analogously, the spin-$\downarrow$ dynamics due to ASEI also influences the tunneling of the hole. Specifically, the presence of the  spin-$\downarrow$ results in a higher hole occupation, which is visible when comparing the dynamics around the black dashed lines in Figs.~\ref{holecomparison}(a) and \ref{holecomparison}(c) and Fig.~\ref{spindowncomparison}(a) and \ref{spindowncomparison}(c) and is seen in detail in Fig.~\ref{holecomparison}(h) where we compare the dynamics of the hole in the right well for $g=2.5$ and $g=5.0$. The black dashed boxes indicate the time intervals for which the hole mainly resides in the right well due to its ASEI dominated tunneling and thus the hole is strongly affected by the presence of the fermion. For stronger $g$ the intervals move to later times due to the $g$ dependence of ASEI. Thus, the increase of the time-scale of the hole dynamics observed in Figs.~\ref{holecomparison}(a) and \ref{holecomparison}(c) are indeed due to the interaction with the spin-$\downarrow$.

These findings indicate correlations of the spin-$\downarrow$ and hole in our system, leading to the absence of the spin-charge separation in the intermediate regime. As we point out in Sec.~\ref{sec:couplingmechanism}, these correlations do not induce magnetic excitations for the strong interaction regime and thereby do not break spin-charge separation. However, the hole remains coupled to the spin-$\downarrow$ particle, which behaves as a non-magnetic impurity. Based on the above, one might be tempted to assume an attractive hole-spin-$\downarrow$ interaction. As will also be discussed in the following section, the precise mechanism leading to the emergence of these correlations is significantly more nuanced.

\subsection{Triplet-Singlet Coupling Mechanism}
\label{sec:couplingmechanism}

 \begin{figure}
    \centering
    \includegraphics[width=1.0\columnwidth]{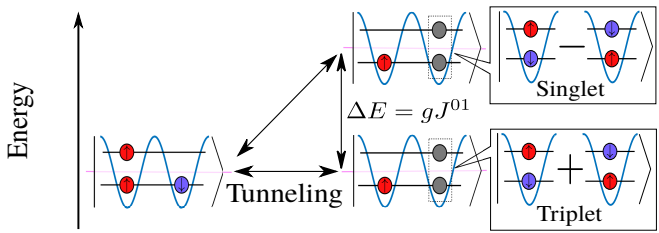}
    \caption{ Schematic illustration of the correlated tunneling process introducing the coupling between the spin and spatial degrees of freedom via the emergence of an interaction-dependent energy gap $\Delta E$ between triplet and singlet states.}
    \label{schematic2}
\end{figure}

Here, we present the mechanism that we hold responsible for the observed dependence of the hole dynamics on the interaction strength $g$. It is based on the occupation of the second energy band which introduces magnetic interactions among the distinct spin-states.

\begin{figure*}
    \centering
    \includegraphics[width=0.95\textwidth]{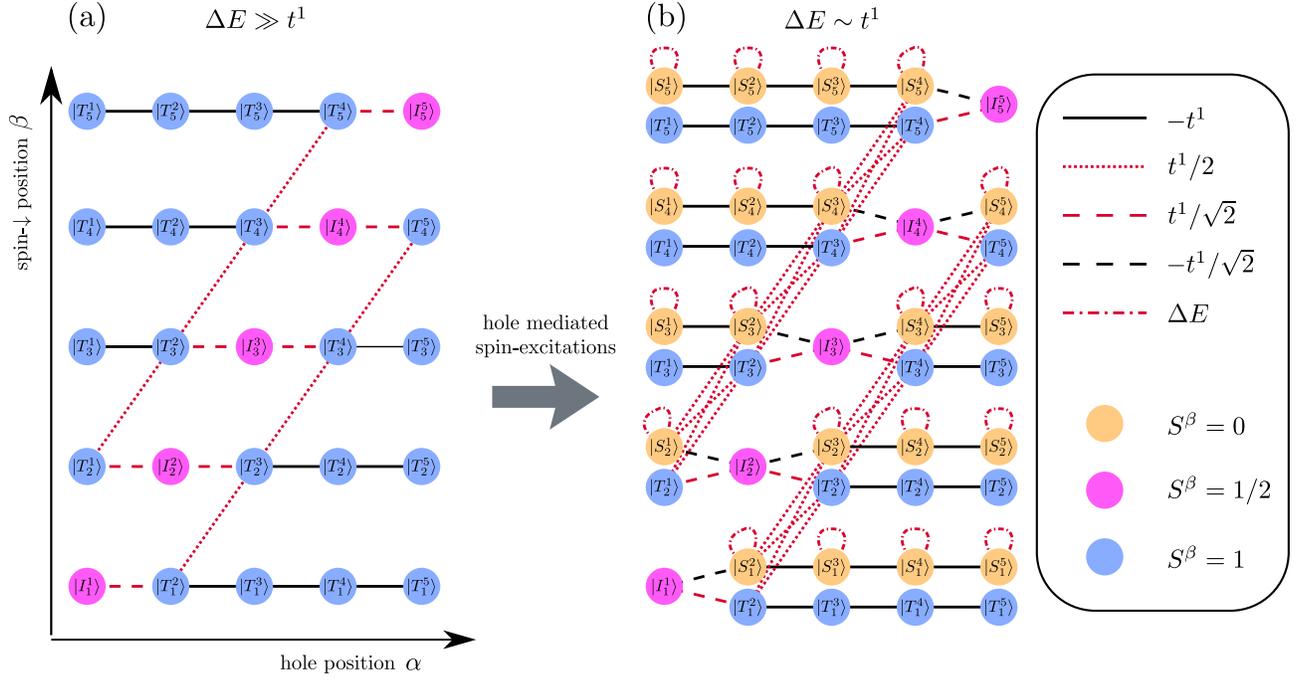}
    \caption{Representation of the matrix elements of Eq.~\eqref{matrix_elements1a}-\eqref{matrix_elements1c} and \eqref{matrix_elements2a}-\eqref{matrix_elements2c} in terms of weighted graphs for the example case of a lattice with five sites. (a) Corresponds to the case of a large gap $\Delta E \gg t^1$, while (b) refers to the general case, where spin-excitations can be induced by the hole dynamics. Each vertex represents one of the states defined in Eq.~\eqref{states_example} and~\eqref{states_example2} with its color representing the value of $S$ at the site where the spin-$\downarrow$ resides (see legend). The edge weights correspond to the values of the corresponding matrix elements which are encoded in the line style, see inset legend.}
    \label{fig:schematic_lattices}
\end{figure*}

Let us begin by considering the initial state [Fig.~\ref{schematic}]. Starting from this configuration, the spin-$\uparrow$ particle occupying the middle well state of the $b=1$ energy band tunnels to the vacant state of the same band in the right well. The thereby created setting where the right well contains one particle of each spin may realize two possible spin configurations, namely a triplet and a singlet state, owing to the spin-dependent interactions \cite{KollerMundinger2015, KollerWall2016, Koutentakis2019}, see Fig.~\ref{schematic2}.

Within the triplet state, the spins of the particles arrange such that their total spin defined by the value of $\hat{S}^2$ is maximal, i.e. they are ferromagnetically ordered. We expect that the interaction energy contribution of this configuration is zero and thus it is degenerate with the initial state, see Fig.~\ref{schematic2}. In contrast, the corresponding singlet state results in a finite interaction energy inducing a total energy increase when compared to the aforementioned states. Thus, for increasing interactions, an energy gap $\Delta E$ develops among the singlet and triplet states, which can be related to ferromagnetic exchange interactions identified in~\cite{KoutentakisMistakidis2020}. When $\Delta E$ becomes much larger than the tunneling energy scale, the singlet states can no longer be accessed when the system is initialized in the state of Eq.~\eqref{initialstate_equ} [Fig.~\ref{schematic}]. Therefore, the system is effectively population trapped in the manifold of triplet states. 

Note that the above argumentation neglects the shift of potential energy due to particle transfer among the sites, which is present due to the broken translational invariance of $V(x)$, but is much smaller than the other system parameters.
In addition, the lowest band tunneling is neglected, a reasonable approximation for small times as it is dominated by ASEI, see Sec.~\ref{sec:spin-down-dynamics}, and thus is much slower than the excited band tunneling. Furthermore, intersite interactions are not considered, since they are negligible for $V_0 = 12 E_{\rm R}$. The validity of these approximations is explicated in Sec.~\ref{sec:effectivemodel_holedynamics} by the introduction of the $tJU$ model.

Below, we employ this mechanism to infer about the presence of spin-charge separation. We first focus on a strict definition of spin-charge separation corresponding to the absence of spin-transport correlations. Subsequently, we comment on a weaker definition in terms of spin-charge deconfinement. A spin-charge deconfined system might possess substantial spin-transport correlations but the involved interactions, are not attractive or strong enough to cause the formation of a bound state of spin and spatial excitations.

For zero spin-spin interactions the spin states of Fig.~\ref{schematic2} are degenerate, which means that the spin-$\uparrow$ particles in the highest band hop independently of one another and so does the hole. The spin-$\downarrow$ particle in this case resides in the lowest band and moves independently from the hole. Thus, spin-charge separation in this case is unambiguous. As we show below, in all other cases it is absent but it can be claimed that within certain time scales it holds {\it approximately}.  In particular, for small but non-zero interactions, resulting in a $\Delta E$ much smaller than the tunneling energy scale, the system behaves similarly to the non-interacting case [see also Fig.~\ref{holecomparison}(a)--\ref{holecomparison}(c)]. However, due to the small energy difference among the singlet and triplet states, a relative phase $e^{-i \Delta E t /\hbar}$  gets accumulated during the dynamics. This results in the {\it dephasing} of the tunneling dynamics on a timescale proportional to $\Delta E$, exhibited as a beating like pattern [see Fig.~\ref{holecomparison}(g)] for $g = 0.5$. This phase accumulation corresponds to the transfer of the spin-$\downarrow$ to the excited band showcasing the development of correlations among the spatial and spin degrees of freedom. Since the development of these correlations is a slow process, one can claim that spin-charge separation approximately holds for short times. 

In order to simplify the description in the case of intermediate interactions, let us first analyze spin-charge separation for strong interactions, where $\Delta E$ is much larger than the tunneling energy splitting. In the latter case, the system is population trapped in the manifold of triplet states and as such spin-charge separation occurs trivially since the hole is not coupled to spin-excitations simply because these do not exist. This results in a system that behaves ferromagnetically (as defined in Ref.~\cite{Koutentakis2019, KoutentakisMistakidis2020}) for all times. However, a different mechanism emerges in this regime, which couples the hole with the spin-$\downarrow$ impurity fermion. To illustrate this, let us consider the tunneling among two adjacent sites $\alpha$ and $\beta$ and the following set of states 
\begin{equation}
\begin{split}
    | I_{\alpha}^{\alpha} \rangle &= \dots \hat{a}^{0 \dagger}_{\alpha \downarrow} \hat{a}^{0 \dagger}_{\beta \uparrow} \hat{a}^{1 \dagger}_{\beta \uparrow} \dots | 0 \rangle, \\
    | T_{\alpha}^{\beta} \rangle &= \dots \hat{a}^{0 \dagger}_{\alpha \uparrow} 
    \frac{\hat{a}^{0 \dagger}_{\beta \uparrow} \hat{a}^{1 \dagger}_{\beta \downarrow} + \hat{a}^{0 \dagger}_{\beta \downarrow} \hat{a}^{1 \dagger}_{\beta \uparrow}}{\sqrt{2}} \dots | 0 \rangle, \\
    | P_{\alpha} \rangle &= \dots \hat{a}^{0 \dagger}_{\alpha \uparrow} \hat{a}^{0 \dagger}_{\beta \uparrow} \hat{a}^{1 \dagger}_{\beta \uparrow} \dots | 0 \rangle, \\
    | I_{\beta}^{\beta} \rangle &= \dots \hat{a}^{0 \dagger}_{\alpha \uparrow} \hat{a}^{1 \dagger}_{\alpha \uparrow} \hat{a}^{0 \dagger}_{\beta \downarrow} \dots | 0 \rangle, \\
    | T_{\beta}^{\alpha}  \rangle  &= \dots  
    \frac{\hat{a}^{0 \dagger}_{\alpha \uparrow} \hat{a}^{1 \dagger}_{\alpha \downarrow} + \hat{a}^{0 \dagger}_{\alpha \downarrow} \hat{a}^{1 \dagger}_{\alpha \uparrow}}{\sqrt{2}} \hat{a}^{0 \dagger}_{\beta \uparrow} \dots | 0 \rangle, \\
    | P_{\beta} \rangle &= \dots \hat{a}^{0 \dagger}_{\alpha \uparrow} \hat{a}^{1 \dagger}_{\alpha \uparrow} \hat{a}^{0 \dagger}_{\beta \uparrow} \dots | 0 \rangle,
\end{split}
\label{states_example}
\end{equation}
where $\hat{a}_{i \sigma}^{b \dagger}$ with spin $\sigma \in \{ \uparrow, \downarrow \}$, site $i \in \{ \alpha, \beta \}$ and the band index $b = 0, 1$ correspond to the fermionic creation operators in the appropriate Wannier state and $| 0 \rangle$ is the fermionic vacuum state. We omit writing explicitly the creation operators for all other sites than $\alpha$ and $\beta$, but we consider them to be the same for all of these states so that the tunneling matrix elements among these states are not trivially zero. In the above introduced notation the position of the hole is given by the state subscript, the position of the spin-$\downarrow$ fermion by the superscript and the value of the total spin in the site that the spin-$\downarrow$ resides by the corresponding letter with $T$ corresponding to $S = 1$ and $I$ to $S = 1/2$. Here, $P$ refers to a spin-$\uparrow$ polarized state and as such has no superscript.
A straightforward calculation of the tunneling matrix elements yields

\setcounter{equation}{7} 
\begin{align}
\label{matrix_elements1a}
    \langle I_{\alpha}^{\alpha} | \sum_{\sigma \in \{ \uparrow, \downarrow \}} -t^1 \left( \hat{a}^{1 \dagger}_{\alpha \sigma} \hat{a}^{1}_{\beta \sigma} + \text{H.c.} \right) | T_{\beta}^{\alpha} \rangle &=  \frac{\sqrt{2} t^1}{2}, \tag{8a} \\
\label{matrix_elements1b}
    \langle T_{\alpha}^{\beta} | \sum_{\sigma \in \{ \uparrow, \downarrow \}} -t^1 \left( \hat{a}^{1 \dagger}_{\alpha \sigma} \hat{a}^{1}_{\beta \sigma} + \text{H.c.} \right) | T_{\beta}^{\alpha} \rangle &=  \frac{t^1}{2}, \tag{8b} \\
\label{matrix_elements1c}
    \langle P_{\alpha} | \sum_{\sigma \in \{ \uparrow, \downarrow \}} -t^1 \left( \hat{a}^{1 \dagger}_{\alpha \sigma} \hat{a}^{1}_{\beta \sigma} + \text{H.c.} \right) | P_{\beta} \rangle &=  -t^1, \tag{8c}
\end{align}
where $t^1$ refers to the tunneling amplitude of band $b=1$. Notice that the mobility of the hole is diminished when it tunnels towards the spin-$\downarrow$ fermion, as the corresponding matrix elements [among $T$ and $I$ states, see Eqs.~\eqref{matrix_elements1a} and ~\eqref{matrix_elements1b}] are smaller in value than when it tunnels among sites containing only spin-$\uparrow$ polarized particles [$P$ states, see Eq.~\eqref{matrix_elements1c}]. We interpret this behavior as an effective repulsion among the spin-$\downarrow$ impurity fermion and the hole, implying the development of correlations among them.

For weaker interactions the singlet states can be populated so we have to consider the influence of the following states in the above effective description, namely,
\begin{equation}
\begin{split}
    | S_{\alpha}^{\beta} \rangle &= \dots \hat{a}^{0 \dagger}_{\alpha \uparrow} 
    \frac{\hat{a}^{0 \dagger}_{\beta \uparrow} \hat{a}^{1 \dagger}_{\beta \downarrow} - \hat{a}^{0 \dagger}_{\beta \downarrow} \hat{a}^{1 \dagger}_{\beta \uparrow}}{\sqrt{2}} \dots | 0 \rangle, \\
    | S_{\beta}^{\alpha} \rangle &= \dots  
    \frac{\hat{a}^{0 \dagger}_{\alpha \uparrow} \hat{a}^{1 \dagger}_{\alpha \downarrow} - \hat{a}^{0 \dagger}_{\alpha\downarrow} \hat{a}^{1 \dagger}_{\alpha\uparrow}}{\sqrt{2}} \hat{a}^{0 \dagger}_{\beta \uparrow} \dots | 0 \rangle,
\end{split} 
\tag{9}
\label{states_example2}
\end{equation}
where $S$ indicates the singlet character of the state corresponding to $S = 0$ within the site where the spin-$\downarrow$ resides.
These states result in the matrix elements
\setcounter{equation}{9}
\begin{align*}
\label{matrix_elements2a}
    \langle I_{\alpha}^{\alpha} | \sum_{\sigma \in \{ \uparrow, \downarrow \}} -t^1 \left( \hat{a}^{1 \dagger}_{\alpha \sigma} \hat{a}^{1}_{\beta \sigma} + \text{H.c.} \right) | S_{\beta}^{\alpha} \rangle &=-\frac{\sqrt{2} t^1}{2},&\tag{10a} \\
\label{matrix_elements2b}
    \langle S_{\alpha}^{\beta} | \sum_{\sigma \in \{ \uparrow, \downarrow \}} -t^1 \left( \hat{a}^{1 \dagger}_{\alpha \sigma} \hat{a}^{1}_{\beta \sigma} + \text{H.c.} \right) | S_{\beta}^{\alpha} \rangle &=\frac{t^1}{2},\tag{10b}\\
\label{matrix_elements2c}
    \langle S_{\alpha}^{\beta} | \sum_{\sigma \in \{ \uparrow, \downarrow \}} -t^1 \left( \hat{a}^{1 \dagger}_{\alpha \sigma} \hat{a}^{1}_{\beta \sigma} + \text{H.c.} \right) | T_{\beta}^{\alpha} \rangle &=\frac{t^1}{2}.\tag{10c}
\end{align*}
They indicate that the hole is effectively repelled by the singlet states similarly to the triplet ones [compare Eqs.~\eqref{matrix_elements1c} and \eqref{matrix_elements2a}--\eqref{matrix_elements2c}]. However, the matrix elements of Eqs.~\eqref{matrix_elements2a}--\eqref{matrix_elements2c} additionally reveal that when the hole is in the same site as the spin-$\downarrow$ or it tunnels towards this site, it can induce magnetic transitions among the triplet and singlet states  [see Eqs.~\eqref{matrix_elements1a} and \eqref{matrix_elements2a}]. Therefore, the coupling mechanism analyzed in the case of strong interactions obtains a magnetic character when $\Delta E$ reduces, resulting in spin-transport correlations. The necessary condition for the occurrence of this spin-transport coupling mechanism is that the energy gap is of the order of the tunneling energy scale of the upper energy band, $\Delta E \sim t^1$. In that case, the gap among the states of Fig.~\ref{schematic2} is large enough such that the dephasing among the triplet and singlet states is as fast as the tunneling of the hole enabling the latter to modify the superposition among these states.

This mechanism reveals that spin-charge deconfinement is present in our system since all of the above-mentioned hole and spin-$\downarrow$ interaction mechanisms are repulsive and no bound state can occur. Therefore, this deconfinement is expected to be observed for an extensive system allowing for sufficient spatial separation of the two (quasi)particles. Even in this case we can show that the correlations stemming from the above-described interaction mechanisms are detectable. To illuminate this, in the following section we develop an effective two-dimensional lattice model that can unravel the correlation mechanisms among the hole and spin-$\downarrow$ particle based on Eqs.~\eqref{matrix_elements1a}--\eqref{matrix_elements1c} and~\eqref{matrix_elements2a}--\eqref{matrix_elements2c}.

\subsection{Hole-spin-impurity correlations within a simplified framework}
\label{sec:couplingmechanism_entanglement}

The matrix elements of Eqs.~\eqref{matrix_elements1a}--\eqref{matrix_elements1c} and \eqref{matrix_elements2a}--\eqref{matrix_elements2c} introduce a very simplistic phenomenological model for the coupling of the hole with spin-$\downarrow$ impurities. In particular, the couplings among the involved states can be represented in terms of the weighted graphs (see also \cite{chartrand1977,trudeau2013}) of Fig.~\ref{fig:schematic_lattices}(a) for $\Delta E \gg t^1$ and Fig.~\ref{fig:schematic_lattices}(b) in case of $\Delta E \sim t^1$. Here each vertex corresponds to one of the states labeled by $|X_{\alpha}^{\beta} \rangle$ with $\alpha, \beta = 1, 2, \dots, M$ referring to the hole and spin-$\downarrow$ position respectively. $M$ is the size of the lattice and $X \in \{ I, S, T \}$ parametrizes the total spin in the site where the spin-$\downarrow$ particle resides [see also Eqs.~\eqref{states_example} and \eqref{states_example2}]. Each edge indicates coupling among these states induced by the tunneling of a particle within the excited band. Notice that all $| S_{\alpha}^{\beta} \rangle$ also have diagonal matrix elements, equal to $\Delta E$, which appear in the graph as closed loops.

The graph of Fig.~\ref{fig:schematic_lattices}(b) depends only on two relevant parameters, the gap $\Delta E$ and the tunneling $t^1$, thus the dynamics of the corresponding Hamiltonian would depend only on the fraction $\Delta E/t^1$ and the initial state. To explore the concepts we have analyzed in Sec.~\ref{sec:couplingmechanism} and to make a connection with the ML-MCTDHX results of Sec.~\ref{sec:holedyn_mlx}, we employ $| \Psi_{\rm gr} (t = 0) \rangle = | I_0^0 \rangle$ as the initial state and consider the state of the system, $| \Psi_{\rm gr} (t) \rangle$, after $t = 10 \hbar/t^1$ in an infinite lattice $\alpha, \beta = 0, \pm 1, \pm 2, \dots$. In practice we employ a finite lattice with $M = 61$ sites which is large enough for convergence to $M \to \infty$ in this time scale within a relative tolerance of $10^{-4}$. 

\begin{figure}
    \includegraphics[width=1.0\columnwidth]{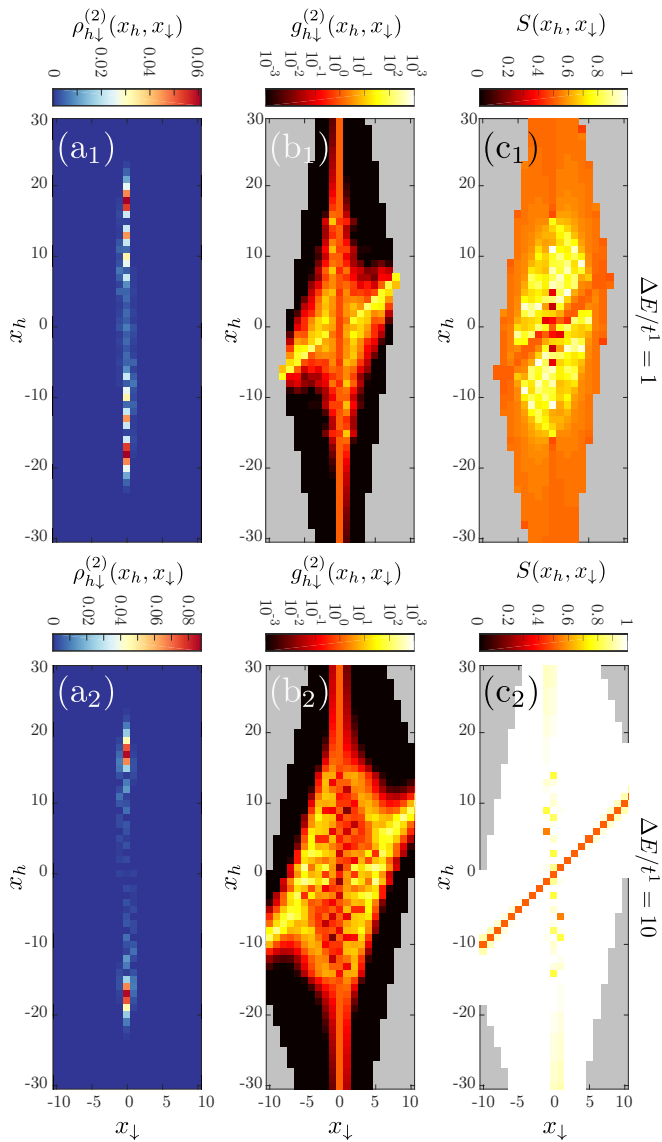}
    \caption{(a$_i$) Two-body density, $\rho^{(2)}_{h \downarrow}(x_h, x_{\downarrow}; t_0)$, and (b$_i$) two-body correlations, $g^{(2)}_{h \downarrow}(x_h, x_{\downarrow}; t_0)$ of a hole and spin-$\downarrow$ (quasi)particles initially localized at $x_{\downarrow} = x_h = 0$ after an expansion time of $t_0 = 10 \hbar/t^1$ for varying value of the singlet-triplet state energy gap ($i = 1$) $\Delta E/t^1 = 1$ and ($i=2$) $\Delta E/t^1 = 10$. (c$_i$) The value of the total spin $S$ in the site of the spin-$\downarrow$ particle as a function of the spin-$\downarrow$ impurity fermion position for the same parameters as (a$_i$) and (b$_i$). The gray regions in (b$_i$) and (c$_i$) indicate the regions where $\rho^{(2)}_{h \downarrow}(x_h, x_{\downarrow}) < (10^{-14})^2$ and thus is numerically indistinguishable from the value zero. In all cases, the Hamiltonian corresponding to Fig.~\ref{fig:schematic_lattices}(b) is employed for a lattice with $M=61$ sites that allows for adequate convergence to the $M \to \infty$ limit.}\vspace{-0.4cm}
    \label{fig:correlations_graphs}
\end{figure}

We are particularly interested in the hole-spin-$\downarrow$ two-body density, $\rho^{(2)}_{h \downarrow}(x_h, x_{\downarrow}; t) = \sum_{X \in \{ I, S, T \}}| \langle X_{x_h}^{x_{\downarrow}} | \Psi_{\rm gr}(t) \rangle |^2$ and the related two-body correlation function
\begin{equation}
    g^{(2)}_{h \downarrow}(x_h, x_{\downarrow}; t) = \frac{\rho^{(2)}_{h \downarrow}(x_h, x_{\downarrow}; t)}{\rho^{(1)}_{h}(x_h; t)\rho^{(1)}_{\downarrow}(x_{\downarrow}; t)},
    \tag{11}
\end{equation}
with $\rho^{(1)}_{h}(x_h; t) = \sum_{X \in \{ I, S, T \}} \sum_{\beta} | \langle X_{x_h}^{\beta} | \Psi_{\rm gr}(t) \rangle |^2$ and $\rho^{(1)}_{\downarrow}(x_{\downarrow}; t) = \sum_{X \in \{ I, S, T \}} \sum_{\alpha} | \langle X_{\alpha}^{x_{\downarrow}} | \Psi_{\rm gr}(t) \rangle |^2$. In addition, the quantity
\begin{equation}
    S(x_h, x_{\downarrow}; t) = \frac{\frac{1}{2} | \langle I_{x_h}^{x_{\downarrow}} | \Psi_{\rm gr}(t) \rangle |^2 +  | \langle T_{x_h}^{x_{\downarrow}} | \Psi_{\rm gr}(t) \rangle |^2}{\rho^{(2)}_{h \downarrow}(x_h, x_{\downarrow})}, 
    \tag{12}
\end{equation}
identifies the average spin $S$ in the site where the spin-$\downarrow$ resides, allowing us to deduce the involvement of magnetic mechanisms in the system.

The results for the hole-spin-$\downarrow$ two-body density after a propagation time of $t = 10 t^1/\hbar$ are provided in Fig.~\ref{fig:correlations_graphs}(a$_1$) and \ref{fig:correlations_graphs}(a$_2$) for $\Delta E/t^1 = 1$ and $\Delta E/t^1 = 10$ respectively. As we have anticipated from the discussion in Sec.~\ref{sec:couplingmechanism}, spin-charge deconfinement is evident in our system. Indeed, the two-body density in Fig.~\ref{fig:correlations_graphs}(a$_1$) and (a$_2$) is large enough to be discernible only in the case that $|x_h| > |x_{\downarrow}|$. Nevertheless, the effect of hole-spin-$\downarrow$ interactions is prominent as increased interactions, reflected by higher values of $\Delta E$, lead to the focusing of the expanding hole probability in a single wavepacket [see Fig.~\ref{fig:correlations_graphs}(a$_2$)] instead of the train of density peaks observed for lower $\Delta E$ [see Fig.~\ref{fig:correlations_graphs}(a$_1$)]. In addition, notice the asymmetric profile of $\rho^{(2)}_{h \downarrow}(x_h, x_{\downarrow}; t)$ with respect to reflection at the $x_{\downarrow}$ axis by inspecting Fig.~\ref{fig:correlations_graphs}(a$_1$) for $-1 < x_{\downarrow} < 1$ and $7 < x_h < 18$ and Fig.~\ref{fig:correlations_graphs}(a$_2$) for $-1 < x_{\downarrow} < 1$ and $0 < x_h < 20$. 
This asymmetry can be attributed to the absence of mirror symmetry of the associated graph, see Fig.~\ref{fig:schematic_lattices}(b), which stems from the fact that the hole has to tunnel to the left in order to push the spin-$\downarrow$ particle to the right and vice versa. This property is better identifiable in the correlation function $g^{(2)}_{h \downarrow}(x_h, x_{\downarrow})$, see Fig.~\ref{fig:correlations_graphs}(b$_1$) and Fig.~\ref{fig:correlations_graphs}(b$_2$). It is shown that the joined probability of finding the hole and spin-$\downarrow$ atom varies by several orders of magnitude compared to the uncorrelated result (notice the logarithmic scale of the color coding in these figures). In particular, the largest values of $g^{(2)}_{h \downarrow}(x_h, x_{\downarrow})$ are observed for $x_h = x_{\downarrow} - 1$ in case of $x_h > 0$ and $x_h = x_{\downarrow} + 1$ for $x_h < 0$. The population of these regions results from the hole shifting the position of the spin-$\downarrow$ fermion repeatedly, substantially delaying the expansion of the latter due to the smaller mobility it has in this case. Nonetheless, since the hole predominantly escapes the region where it interacts with the spin-$\downarrow$, the probability of lying there given by $\rho_{h \downarrow}^{(2)}(x_h, x_{\downarrow})$ is very small [Fig.~\ref{fig:correlations_graphs}(a$_1$),Fig.~\ref{fig:correlations_graphs}(a$_2$)].

Finally, Fig.~\ref{fig:correlations_graphs}(c$_1$) and Fig.~\ref{fig:correlations_graphs}(c$_2$) reveal our arguments regarding spin-charge separation. For $\Delta E/t^1 = 10$, we observe that the configurations of maximal possible spin, i.e. $S = 1/2$ for $x_h = x_{\downarrow}$ and $S = 1$ for $x_{\downarrow} \neq x_h$, are almost exclusively populated due to the large gap among the triplet and singlet configurations. Nevertheless, some spin-excited sites exist corresponding predominantly to the additional expanding wave packets of the hole-density at $-1 < x_{\downarrow} < 1$ and $7 < | x_h | < 14$ [see Figs.~\ref{fig:correlations_graphs}(c$_2$) and \ref{fig:correlations_graphs}(a$_2$)]. Notice that for the maximal value $S = 1$ the spin-$\downarrow$ does not correspond to a spin-excitation since ferromagnetism is controlled by the total spin rather than polarization~\cite{Koutentakis2019, KoutentakisMistakidis2020}. Thus, within a reasonable approximation, we can claim that no substantial correlations among the spin-excitations and the hole transport develop in this regime while the hole and spin-$\downarrow$ (acting here as a non magnetic impurity) are strongly coupled [see Fig.~\ref{fig:correlations_graphs}(b$_2$)].
For smaller gaps $\Delta E = t^1$, the magnetic response of the system is substantially different. In particular, we can see in Fig.~\ref{fig:correlations_graphs}(c$_1$) that high $S > 0.7$ is realized only in the region of $| x_h | < 15$ while outside of this region $S \approx 0.5$, corresponding to the state that the spin-$\downarrow$ remains in the lowest band of the lattice. Importantly, the different peaks in the train of the hole density that is expelled from the site $0$ [Fig.~\ref{fig:correlations_graphs}(a$_1$) for $7 < x_h < 23$ and $x_{\downarrow} = 0$] correspond to different values of $S$ [see the variation of $S$ in Fig.~\ref{fig:correlations_graphs}(c$_1$) when $x_h$ increases from $x_h = 7$ to $23$]. This illustrates the long-range entanglement of spin-excitations and hole position in this setup, which can be utilized to bring the spin-$\downarrow$ impurity fermion to the desired $S$ state by manipulating the hole position that lies far away from it. This property might be useful for spintronic applications~\cite{vanderStraten2013}. Thus, we can claim that the spin and spatial states of our setup are coupled despite the fact that spin-charge deconfinement is exhibited.

\section{Effective Spin-Chain Model}
\label{sec:effectivemodel}
Here we introduce an effective model that provides the connecting link among the phenomenological description, that identified spin-charge correlations, and the results obtained with ML-MCTDHX, which indicated spin-charge separation. It rigorously reveals the emerging energy gap between singlet and triplet states, upon which the phenomenological arguments of Sec.~\ref{sec:couplingmechanism} are based, in terms of the microscopic system parameters. This enables the identification and assignment of the described phenomenology in the {\it ab initio} results. 

\subsection{Effective Hamiltonian}
\label{sec:effectivemodel_hamilton}
We introduce an effective $tJU$ model generalizing the approach used in \cite{KoutentakisMistakidis2020}. Within this model a single site refers to the position on a band $b$ inside a specific well $s$. The sites are well-captured by employing the previously mentioned Wannier states $\phi_s^b(x)$ as a basis set, in which a fermion of spin-$\sigma$ is created by using the operator $\hat{a}^{b \dagger}_{s \sigma}$. 

Let us first introduce the non-interacting part of the effective Hamiltonian which reads as~\cite{KoutentakisMistakidis2020}
\begin{equation}
    \begin{split}
      \hat{H}_0 = -\sum_{b=0,1} \sum_{\sigma \in \{\uparrow,\downarrow\}} t^b \left( 
        \hat{a}^{b \dagger}_{L\sigma} \hat{a}^{b}_{M\sigma} 
        + \hat{a}^{b \dagger}_{M\sigma} \hat{a}^{b}_{R\sigma} 
        + \text{H.c.} \right) \\ 
      + \sum_{b=0,1} \sum_{\sigma \in \{\uparrow,\downarrow\}}\bigg[(\epsilon^b +\Delta\epsilon^b)(\hat{n}^{b}_{L\sigma} +\hat{n}^{b}_{R\sigma} ) + \epsilon^b \hat{n}^{b}_{M\sigma}\bigg],
    \end{split}
    \label{effectiveH_non_interact_equ}
    \tag{13}
\end{equation}
(see Appendix \ref{sec:non-interacting-effective}). The first part of the Hamiltonian describes the single particle tunneling among the wells. In each band $b$, we consider only next-nearest neighbor tunneling which occurs with a tunneling amplitude $t^b$. The second term describes the energy of non-interacting particles, where $\epsilon ^b$ is the average energy of the eigenstates that form the band $b$. As pointed out in Sec.\ref{sec:holedyn_mlx}, the translational invariance of the lattice is broken and the left and right wells acquire an additional energy offset, $\Delta \epsilon ^b$, when compared to the central well. Notice that this offset is identical for the left and the right wells due to the parity symmetry $x \to -x$ of $V(x)$.

The exact description including interactions within the tight binding model of Eq.~\eqref{effectiveH_non_interact_equ}, considers all possible matrix elements between different Wannier states. Such a framework is undesirable for us as it would obscure the interpretation of the system dynamics. Therefore, we use a minimal effective description, where we restrict the model to on-site interactions. This is a valid approximation assuming that the Wannier states are well localized in each well, which is adequate for the large values $V_0$ employed here. These types of Fermi-Hubbard models have been successfully applied to fermions in double wells and lattice traps \cite{KoutentakisMistakidis2020,Lewenstein2007,Esslinger2010}. Here, we consider intraband interactions with strength $U^b$ and interband interactions $J^{b b^\prime}$ in each well. These interaction constants can be derived from the Wannier states: $J_s^{b b^\prime} = \int \mathrm{d}x |\phi_s^b(x)|^2 |\phi_s^{b^\prime}(x)|^2 $, $U_s^b  =J_s^{b b} = \int \mathrm{d}x |\phi_s^b(x)|^4 $. Notice the lattice dependence $s \in \{\mathrm{L}, \mathrm{M}, \mathrm{R} \}$ of the above interaction constants due to the broken translational invariance.

To account for interactions, we introduce the following Hamiltonian that consists of direct interaction terms:
\begin{equation}
    \begin{split}
        \hat{H}_{I,\text{dir}} =& g \left[ \sum_{b=0,1}\sum_{s\in\{\mathrm{L}, \mathrm{M},\mathrm{R}\}} U^b_s \hat{n}^{b}_{s\uparrow}\hat{n}^{b}_{s\downarrow} \right. \\
        &+ \left. \sum_{s\in\{\mathrm{L}, \mathrm{M},\mathrm{R}\}} J_s^{01} (\hat{n}^{0}_{s\uparrow}\hat{n}^{1}_{s\downarrow} + \hat{n}^{1}_{s\uparrow}\hat{n}^{0}_{s\downarrow} )\right].
    \end{split}
    \label{effectiveH_interact_equ_direct}
    \tag{14}
\end{equation}
This Hamiltonian includes interactions by counting the particles of both spin species in each well and assigning them with the corresponding interaction strength. Additionally, we include exchange interaction terms of the form $-g\,J^{01}_s(\hat{a}^{0\dagger}_{s\sigma}\hat{a}^{1 \dagger}_{s\sigma^\prime}\hat{a}^{1}_{s\sigma}\hat{a}^{0}_{s\sigma^\prime} )$ for all different combinations $\{s,\sigma,\sigma^\prime\}$. They describe the process where two fermions located inside the same well but in different bands can exchange their spin. By adding these to $\hat{H}_{I,\text{dir}}$, we can write the complete effective interaction Hamiltonian as
\begin{equation}
    \begin{split}
        \hat{H}^\prime_I =&g \left[ \sum_{b=0,1}\sum_{s\in\{\mathrm{L}, \mathrm{M},\mathrm{R}\}} U^b_s \hat{n}^{b}_{s\uparrow}\hat{n}^{b}_{s\downarrow} \right. \\
        &- \left. \sum_{s\in\{\mathrm{L}, \mathrm{M},\mathrm{R}\}}J_s^{01} \left(\hat{\mathbf{S}}_s^0\cdot \hat{\mathbf{S}}_s^1 -\frac{1}{4} \hat{n}^{0}_{s} \hat{n}^{1}_{s}\right) \right],
    \end{split}
    \label{effectiveH_interact_equ}
    \tag{15}
\end{equation}
where $ \hat{n}^{b}_{s} =\hat{n}^{b}_{s \uparrow} +\hat{n}^{b}_{s \downarrow}$ counts the total number of particles inside a well. Above, the operators $\hat{\mathbf{S}}_s^b$ are the spin operators $\hat{\mathbf{S}}_s^b =\hat{S}_{x,s}^b \mathbf{\hat{e}_i}+ \hat{S}_{y,s}^b \mathbf{\hat{e}_j} +\hat{S}_{z,s}^b \mathbf{\hat{e}_k}$ where $\mathbf{\hat{e}_i} $ are the unit vectors in spin space and $ \hat{S}^b_{i,s}= \frac{1}{2}\sum_{\sigma \sigma^\prime} \pmb{\sigma}^i_{\sigma \sigma^\prime} \hat{a}^{b \dagger}_{s\sigma} \hat{a}^{b}_{s\sigma^\prime}$ with $i \in \{x,y,z\}$, $s \in \{ \mathrm{L},\mathrm{M},\mathrm{R}\}$ and $\pmb{\sigma}^i$ being the Pauli matrices. Finally, the complete effective $tJU$ model is given by: $\hat{H}_{tJU}= \hat{H}_0 + \hat{H}^\prime_I$. 
This model splits the Hilbert space of the total system into different subspaces corresponding to distinct band configurations $\{ n^0 = \sum_s n^0_s, n^1 = \sum_s n^1_s,\dots \}$ (see also~\cite{KollerMundinger2015, Koutentakis2019}) that do not couple with one another. This approximation is well-founded since $\epsilon^b$ is the largest energy scale of the system. For more details on this approach, see Appendix~\ref{app:effective}.

\subsection{Comparison with the {\it ab initio} dynamics}
\label{sec:effectivemodel_holedynamics}
As a first probe of the effective model, we compare its hole dynamics [see Figs.~\ref{holecomparison}(d)--\ref{holecomparison}(f)] to the one obtained from ML-MCTDHX [see Figs.~\ref{holecomparison}(a)--(\ref{holecomparison}c)]. Within $tJU$, the system is initialized in the state of Eq.~\eqref{initialstate_equ} while its time evolution follows $| \Psi(t) \rangle = \exp (-i \hat{H}_{tJU} t/\hbar ) | \Psi(0) \rangle$. The position of the hole [Figs.~\ref{holecomparison}(d)--\ref{holecomparison}(f)] can be tracked exactly by the expectation values of the hole operators
\begin{equation}
\begin{split}
\tilde{h}_s(t) &= \langle \Psi (t) | \left( 1 - \hat{a}^{1 \dagger}_{s \uparrow}\hat{a}^{1}_{s \uparrow} \right)\left( 1 - \hat{a}^{1 \dagger}_{s \downarrow}\hat{a}^{1}_{s \downarrow} \right) | \Psi (t) \rangle, \\
&= 1 - \langle \Psi (t) | \hat{n}^{1 \dagger}_{s} | \Psi (t) \rangle + \langle \Psi (t) | \hat{n}^{1 \dagger}_{s \uparrow} \hat{n}^{1 \dagger}_{s \downarrow} | \Psi (t) \rangle, \\
\end{split}
\tag{16}
\label{hole_operator_tJU}
\end{equation}
with $s \in \{ L, M, R \}$. Note that $\tilde{h}_s(t)$ is much more difficult to experimentally measure than $h_s(t)$ since it requires that the distinct bands of the lattice can be resolved individually.
The quantity $h_s(t)$ circumvents that by adding the negligible contribution of the first band $\delta h_s = 1 - \langle \Psi (t) | \hat{n}^{0 \dagger}_{s} | \Psi (t) \rangle + \langle \Psi (t) | \hat{n}^{0 \dagger}_{s \uparrow} \hat{n}^{0 \dagger}_{s \downarrow} | \Psi (t) \rangle \approx 0$ provided that no doublons form (see also Appendix~\ref{app:doublons}). In the following we will directly compare $\tilde{h}_s(t)$ and $h_s(t)$ despite that $\tilde{h}_s(t) \neq h_s(t)$ in the general case.

Indeed, Figs.~\ref{holecomparison}(d)--\ref{holecomparison}(f) reveal that the main qualitative structures of the {\it ab initio} approach are reproduced within the $tJU$ model.
In particular, we can identify the same three interaction regimes that partition the dynamics obtained from ML-MCTDHX: weak ($0.3 \leq g$), intermediate ($0.3 < g < 1.0 $), and strong interactions ($1.0\leq g$) [compare, e.g., Figs.~\ref{holecomparison}(a) and \ref{holecomparison}(d)]. Even more importantly the qualitative behavior within these regimes is almost equivalent to the {\it ab initio} case (see also Sec~\ref{sec:holedyn_mlx}).
The tunneling dynamics mainly involves the left and right wells, which is consistent with the ML-MCTDHX case [see Figs.~\ref{holecomparison}(a)--\ref{holecomparison}(c)] and is attributed to the energetic separation of the central well due to $\Delta \epsilon^b$ [see Eq.~\eqref{effectiveH_non_interact_equ}]. The above also indicates the relevance of our findings as the experimentally accessible $h_s(t)$ is proven to be an adequate probe of the spin-charge separation in our system, even when compared with the theoretically more rigorous measure $\tilde{h}_s(t)$ (see also Appendix~\ref{app:doublons}).

Let us now compare the particle dynamics for the spin-$\downarrow$ particle. To this end, we show the particle-number dynamics $n_{s \downarrow}(t)$ of Eq.~\eqref{spin_down_occ} in Fig.~\ref{spindowncomparison}. It can be observed that the $tJU$ model qualitatively captures the interaction dependence of the spin-$\downarrow$ dynamics [compare Figs.~\ref{spindowncomparison}(d) and \ref{spindowncomparison}(f)] which, as claimed in Sec.~\ref{sec:spin-down-dynamics}, stems from the ASEI. In particular, the corresponding coupling of this interaction mechanism within the $tJU$ approach is $\sim 3(t^1)^2/(g \bar{U}^1)$, which can be shown to well describe the spin-$\downarrow$ dynamics for $g > 3$ (not shown here for brevity). In addition, the scattering events among the hole and spin-$\downarrow$ (quasi)-particles are also clearly captured by the $tJU$ model. Finally, the accumulation of hole density within the site where the spin-$\downarrow$ resides is observable [see Figs.~\ref{holecomparison}(d) and \ref{holecomparison}(f)]. This is attributed to the reduced mobility of the hole when placed in the vicinity of the spin-$\downarrow$ particle [see Eqs.~\eqref{matrix_elements1a},\eqref{matrix_elements1b}, and~\eqref{matrix_elements2a}--\eqref{matrix_elements2c}], leading to the accumulation of its density in the site of the impurity.

Nevertheless, there are two notable differences between the effective $tJU$ model and the ML-MCTDHX results. First, the involved time scales slightly differ, which can be seen by comparing the time scales of the ASEI [compare the dynamics around the black dashed lines in Figs.~\ref{spindowncomparison}(a)--\ref{spindowncomparison}(c) to \ref{spindowncomparison}(d)--\ref{spindowncomparison}(f)] and also by studying the hole dynamics [see Fig.~\ref{holecomparison}(a)--\ref{holecomparison}(c) and \ref{holecomparison}(d)--\ref{holecomparison}(f)]. This frequency difference can be attributed to the increased level repulsion within the eigenspectrum in the {\it ab initio} case, stemming from the terms neglected within the $tJU$ approximation [see also Sec.~\ref{sec:spectrum}]. Second, the effective model is missing the weak resonances at $g\simeq3.0$ and $g \simeq 3.8$. This discrepancy is visible for both the hole (Fig.~\ref{holecomparison}) and the spin-$\downarrow$ dynamics (Fig.~\ref{spindowncomparison}). As we argue in Sec.~\ref{sec:spectrum} these resonances can be attributed to an interband transfer process.

\subsection{Emergence of the Triplet-Singlet Energy Gap}
\label{sec:effective_Egap}
We now proceed to use the $tJU$ model to unravel the coupling mechanism between singlet and triplet states (see Fig.~\ref{schematic2}) that we claim to be responsible for the development of spin-charge correlations in our system.

\begin{figure}
    \centering
    \includegraphics[width=1.0\columnwidth]{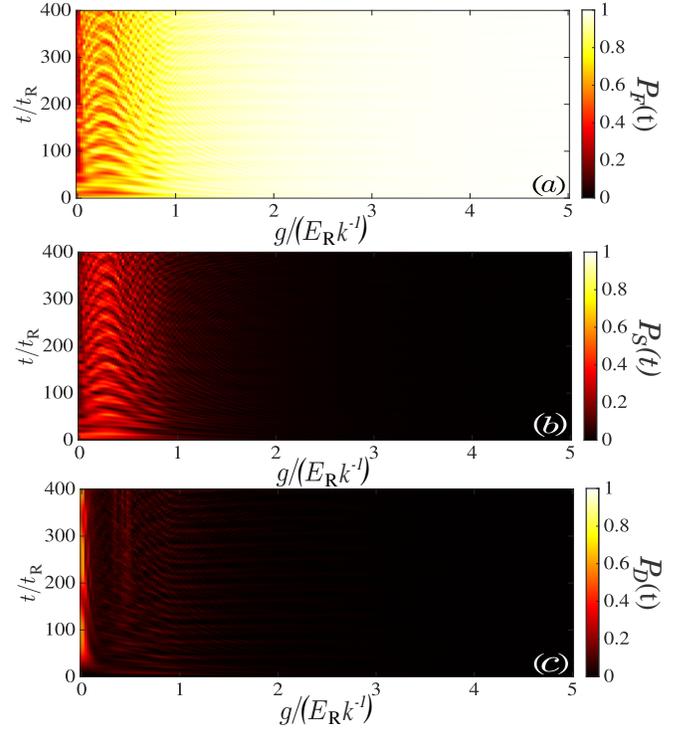}
    \caption{ (a) Probability to occupy the ferromagnetically ordered triplet and isolated spin-$\downarrow$ states, $P_F(t)$, (b) the same for the singlet states, $P_S(t)$, and (c) for doublon states, $P_D(t)$, within the effective $tJU$ model.}
    \label{tripletsinglets}
\end{figure}

We consider a well $s$ containing one spin-$\uparrow$ and one spin-$\downarrow$ particle and examine the interaction part of the effective Hamiltonian [see Eq.~\eqref{effectiveH_interact_equ}]. The first part $\propto U^b_s$, which considers direct on-site interactions between both spin species, does not contribute here except for very small interactions, since doublon formation is negligible $\langle \Psi(t) | \hat{n}^{b}_{s\uparrow}\hat{n}^{b}_{s\downarrow} | \Psi(t) \rangle \approx 0$ (see Appendix~\ref{app:doublons}).
In contrast, the second part of the interaction Hamiltonian $\propto J_s^{b b'}$ [see Eq.~\eqref{effectiveH_interact_equ}], which describes interband interactions, acquires different values depending on the spin configuration. Let us first consider a spin triplet state. The spin operators yield $\langle T_{\alpha}^{\beta} | \hat{\mathbf{S}}_s^0\cdot \hat{\mathbf{S}}_s^1 | T_{\alpha}^{\beta} \rangle = 1/4$ for $\alpha \neq s$ and all $\beta$. Additionally, $\langle T_{s}^{\beta} | \hat{\mathbf{S}}_s^0\cdot \hat{\mathbf{S}}_s^1 | T_{s}^{\beta} \rangle = 0$ holds for all $\beta$ which results in an interaction energy of $E^T_{I} =\langle \hat{H}_I\rangle^T= 0$. Similar results are also reproduced for the $| I_{\alpha}^{\alpha} \rangle$ states. On the other hand, for the singlet state $\langle S_{\alpha}^s | \hat{\mathbf{S}}_s^0\cdot \hat{\mathbf{S}}_s^1 | S_{\alpha}^s \rangle = - 3/4$, $\langle S_s^{\beta} | \hat{\mathbf{S}}_s^0\cdot \hat{\mathbf{S}}_s^1 | S_s^{\beta} \rangle = 0$ and $\langle S_{\alpha}^{\beta} | \hat{\mathbf{S}}_s^0\cdot \hat{\mathbf{S}}_s^1 | S_{\alpha}^{\beta} \rangle = 1/4$ for $\alpha \neq s \neq \beta$, yielding $E^S_{I} =\langle \hat{H}_I\rangle^S= g J_s^{01}$. The energy difference between both spin configurations is $\Delta E = E^S_{I} -E^T_{I} = g J_s^{01}$. Thus, this gap is indeed interaction-dependent as we have previously claimed in Sec.~\ref{sec:couplingmechanism}. The proportionality is characterized by the interband interaction constant $J_s^{01}$, that depends on the well $s$. However, $J_M^{01} \approx J_L^{01} = J_R^{01} \equiv J^{01}$ and thus we expect the same interaction dependent behavior for all wells.

The interaction energy of the initial state inside the right well, where the spin-$\downarrow$ particle is isolated, reads $E^{0}_{I} =\langle \hat{H}_I\rangle^0 = 0 = E^T_{I}$, which is the same as for the triplet configuration. The relative energies for different spin configurations are illustrated in Fig.~\ref{schematic2}. Since triplet states are not energetically cut off from the initial state, they are anticipated to take part in the dynamics for all interaction strengths. Singlet states are expected to contribute only up to a critical interaction strength, below which the energy gap $\Delta E = g J^{01}$ is small enough to be bridged, similarly to our arguments in Sec.~\ref{sec:couplingmechanism}.

In order to check this hypothesis, we evaluate the time-dependent probability for the system to be in a ferromagnetically ordered state, i.e. $| T_{\alpha}^{\beta} \rangle$ and $| I_{\alpha}^{\alpha} \rangle$ as in Eq.~\eqref{states_example},
\begin{equation}
\begin{split}
    P_F(t) =& \sum_{\alpha \neq \beta \in \{ {\rm L}, {\rm M}, {\rm R} \}} | \langle T_{\alpha}^{\beta} | \Psi(t) \rangle |^2  \\
    & + \sum_{\alpha \in \{ {\rm L}, {\rm M}, {\rm R} \}} | \langle I_{\alpha}^{\alpha} | \Psi(t) \rangle |^2.
\end{split}
\tag{17}
\end{equation}
Analogously, the corresponding probability for the singlet states, see also Eq.~\eqref{states_example2}, is defined by
\begin{equation}
\begin{split}
    P_S(t) =\sum_{\alpha \neq \beta \in \{ {\rm L}, {\rm M}, {\rm R} \}} | \langle S_{\alpha}^{\beta} | \Psi(t) \rangle |^2.
\end{split}
\tag{18}
\end{equation}
The results of these evaluations are presented in Fig.~\ref{tripletsinglets}(a) and (b). This figure indicates that the ferromagnetically ordered states participate in the dynamics for the whole range of interactions, see Fig.~\ref{tripletsinglets}(a), and are the dominant contribution to $| \Psi(t) \rangle$ for strong interactions $g \geq 1.0$. In this regime, singlets do not contribute to the dynamics as indicated by the vanishing overlap probability, see Fig.~\ref{tripletsinglets}(b). For interactions below that value, $g \leq 1.0$, $P_F(t)$ and $P_S(t)$ show that triplet and singlet states both take part equally in the dynamics. Additionally, there is an apparent coupling between triplet and singlet states inside the intermediate regime, $0.3 \leq g \leq 1.0$, identified by the visible oscillatory behavior of $P_F(t)$ and $P_S(t)$. Since this is the regime for which we have observed broken spin-charge separation (see also Sec.~\ref{sec:holedyn_mlx}), this strengthens our hypothesis that this coupling mechanism is in fact responsible for the presence of spin-charge correlations and the consequent absence of spin-charge separation in our system. For completeness, note that for $g=0$ triplets and singlets both have a less important role in the dynamics [see Figs.~\ref{tripletsinglets}(a) and \ref{tripletsinglets}(b) for $g < 0.1$]. This occurs because in this case doublon states are present that were negligible for larger interactions [Fig.~\ref{tripletsinglets}(c)] (see also Appendix~\ref{app:doublons}). Owing to the relation $P_D(t) = 1 - P_F(t) - P_S(t)$, that holds both within the $tJU$ model as well as in the {\it ab initio} system, they appear as simultaneous depletion in both $P_S(t)$ and $P_T(t)$.

\subsection{Effective Eigenspectrum}
\label{sec:spectrum}

\begin{figure}
    \centering
    \includegraphics[width=1.0\columnwidth]{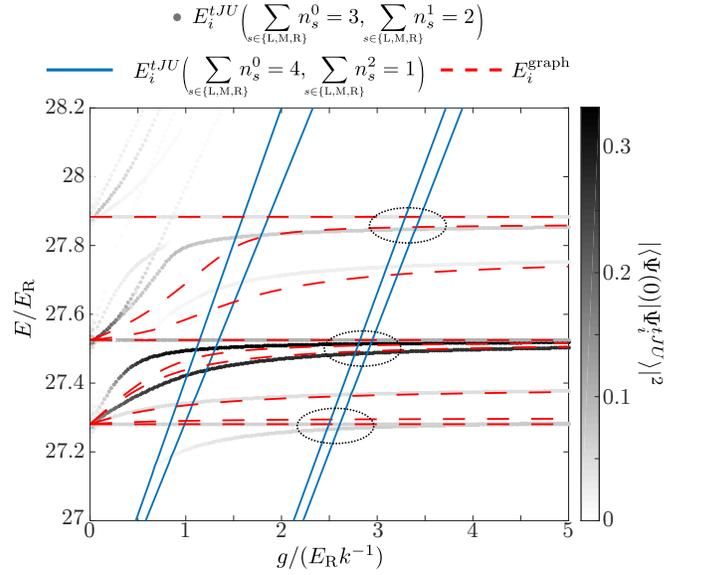}
   \caption{The eigenspectrum of the $tJU$ model for $N_{\uparrow} = 4$ and $N_{\downarrow} = 1$ fermions confined in a triple well with $V_0 = 12 E_{\rm R}$. The overlap of the depicted eigenstates with the initial state $|\langle \Psi(0) | \Psi_i^{tJU} \rangle |^2$ is indicated by the grayscale of the data points. The dashed lines indicate the nine energetically lowest eigenstates of the graph Hamiltonian, Fig.~\ref{fig:schematic_lattices}(b). The solid blue lines indicate the $tJU$ eigenstates with $4$ and $1$ particles in the $b=0$ and bands, respectively. These states result in the resonances observed within ML-MCTDHX in the $g$ regions indicated by the dotted ellipses due to their beyond $tJU$ couplings to the manifold of relevant states (see text).}
    \label{spectrum}
\end{figure}

In this section, we consider the spectrum of the effective $tJU$ model (see Fig.~\ref{spectrum}) in order to compare with the phenomenological model of Sec.~\ref{sec:couplingmechanism_entanglement} and to explain the missing features present in the \textit{ab initio} many-body results. We focus on the eigenstates possessing significant overlap with the initial state of Eq.~\eqref{initialstate_equ}. The spectrum reveals that the initial state involves the occupation of a multitude of eigenstates for all values of $g$, which is expected as hole dynamics is observed in all considered cases [see Figs.~\ref{holecomparison}(d)--\ref{holecomparison}(f)].

To get a deeper understanding of the above occupation pattern of the $tJU$ eigenspectrum, we compare it with the relevant nine energetically lowest states of the graph of Fig.~\ref{fig:schematic_lattices}(b), appearing as dashed lines in Fig.~\ref{spectrum}. In order to facilitate a comparison between both spectra, we include the specific site dependence of the triplet-singlet gap stemming from the variation of $J^{01}_s$ for different $s \in \{\mathrm{L}, \mathrm{M}, \mathrm{R}\}$. Furthermore, the energetic barrier of the spin-$\downarrow$ particle to occupy the middle site, stemming from the expulsion of the $b=1$ particles to the side wells, is also taken into account.
This leads to the modification of diagonal matrix elements of the graph Hamiltonian to
\begin{equation}
\begin{split}
    \langle I_{\alpha}^{\alpha} | \hat{H}_{\rm graph} | I_{\alpha}^{\alpha} \rangle &= E_0 + \delta_{\alpha M} \Delta \epsilon^1, \\
    \langle T_{\alpha}^{\beta} | \hat{H}_{\rm graph} | T_{\alpha}^{\beta} \rangle &= E_0 +\delta_{\alpha M} \Delta \epsilon^1, \\
    \langle S_{\alpha}^{\beta} | \hat{H}_{\rm graph} | S_{\alpha}^{\beta} \rangle &= E_0 + \Delta E^{\beta} + \delta_{\alpha M} \Delta \epsilon^1 \\
    &= E_0 + g J^{0 1}_{\beta}+ \delta_{\alpha M} \Delta \epsilon^1,
\end{split}
\tag{19}
\end{equation}
where $\alpha, \beta \in \{ {\rm L}, {\rm M}, {\rm R} \}$ and $E_0 = 3 \epsilon^0 + 2 \Delta \epsilon^0 + 2 \epsilon^1 + \Delta \epsilon^1$ is the appropriate zero energy offset such that the spectra of both approaches agree for $g=0$. The remaining matrix elements are given by Eqs.~\eqref{matrix_elements1a}--\eqref{matrix_elements1c} and \eqref{matrix_elements2a}--\eqref{matrix_elements2c}.

With these corrections incorporated, we observe that the spectra of both effective models show excellent agreement for strong interactions, $g>2$, indicating the tendency of the system to provide a spin-$\downarrow$ and hole entangled state, as unravelled in Secs.~\ref{sec:couplingmechanism} and~\ref{sec:couplingmechanism_entanglement}. 
For weaker interactions the agreement between the methods is only qualitative, thus additional mechanisms to the ones captured within $\hat{H}_{\rm graph}$ are involved in the dynamics.
In particular, $\hat{H}_{\rm graph}$ completely neglects the presence of doublon states.
Although these are not significantly occupied within the $tJU$ model [see Fig.~\ref{tripletsinglets}(c)], they have the important role of increasing the spin-$\downarrow$ mobility via the ASEI mechanism~\cite{Anderson1959, MazurenkoChiu2017}. This significantly modifies the behavior of the $tJU$ eigenspectrum in the regime $0.5 < g < 2$.
Finally, for $g < 0.5$ a multitude of avoided crossings among the doublon states and the remaining configurations are observed explaining their involvement within the weakly interacting regime (see Sec.~\ref{sec:effective_Egap}).

The additional resonances observed within the ML-MCTDHX approach can also be explained in terms of Fig.~\ref{spectrum}. The solid blue lines indicate the $tJU$ eigenenergies corresponding to states where four particles are occupying the ground band $b=0$ and one particle lies in the second excited band $b = 2$. These configurations possess a smaller single-particle energy than the states with three and two particles in the $b = 0$ and $1$ bands, respectively, which we have been considered up to now. Due to the necessity to generate a ground band doublon in the former band configuration, the energies of these states increase significantly $\propto g \bar{U}^0$ for increasing repulsion. As a consequence, they cross the manifold of the considered states for a finite value of $g$, as it can be seen in Fig.~\ref{spectrum} for $2 < g < 4$ range, see dotted ellipses. For smaller $g \approx 1$, the crossings observed in Fig.~\ref{spectrum} give rise to very narrow resonances that we cannot resolve in our {\it ab initio} calculations. These crossings are exact within the $tJU$ description because the related interaction terms, that result in the coupling of those distinct band configurations, are neglected \cite{KollerWall2016, Koutentakis2019, KoutentakisMistakidis2020}. In particular, these terms correspond to the so-called cradle mode \cite{MistakidisCao2014, MistakidisCao2015, MistakidisKoutentakis2018}, where two particles of the same band are expelled to different bands due to interaction.

\section{Conclusions and Outlook}
\label{sec:conclusion}
We have investigated the correlated dynamics of a spin-$\frac{1}{2}$ fermionic system confined in a one-dimensional triple-well potential. The considered initial state is composed of a polarized ensemble of spin-$\uparrow$ fermions confined in a double-well subsystem and a single spin-$\downarrow$ fermion trapped in the remaining well of the overall potential. By implementing this initial state with a vacancy in the rightmost well, we are able to probe the presence of the usually assumed spin-charge separation both in terms of development of spin-transport correlations and spin-charge deconfinement. By exploring the hole dynamics for different interaction strengths among the spin species, we have revealed the development of spin-charge correlations in this multi-band setup. In particular, in the intermediate interaction regime, where neither interband exchange interaction nor the development of magnetic excitations can be neglected, the system does not exhibit spin-charge separation since strong correlations of the spin and hole dynamics develop. Spin-charge separation is re-established when leaving this regime to higher or lower interactions since one of the above mentioned supporting mechanisms becomes negligible in either case. The mechanism that couples spin triplet and singlet configurations into an effective two-dimensional lattice has been elucidated in terms of a phenomenological description. The latter is based on a graph generated from the tunneling-induced couplings among the involved states. This simple description demonstrates the emergence of spin-$\downarrow$ and hole entanglement and the relation to magnetic excitations in the intermediate interaction regime. We have characterized this mechanism by comparisons among this phenomenological description, the coarse-grained effective $tJU$ model and the {\it ab initio} ML-MCTDHX approach, demonstrating the persistence of the development of spin-$\downarrow$ and hole correlations when the system is treated within different levels of rigor.

This work sets only the beginning of studying the absence of spin-charge separation in one-dimensional multi-band systems. An interesting direction is the study of larger systems with more than one spin-$\downarrow$ particle that enable the examination of more complex spin-exchange dynamics. Furthermore, the presence of an additional hole in a larger system gives rise to the possibility of hole-hole interactions and may also lead to interactions of the 1D-equivalent of magnetic polarons. 

\begin{acknowledgements}
This work has been funded by the Cluster of Excellence
``Advanced Imaging of Matter'' of the Deutsche Forschungsgemeinschaft
(DFG) - EXC 2056 - project ID 390715994. G. M. K. gratefully acknowledges funding from the European Union’s Horizon 2020 research and innovation programme under the Marie Skłodowska-Curie grant agreement No.~ 101034413.
\end{acknowledgements}

\appendix
\section{ML-MCTDHX}
\label{sec:mlx}

In order to obtain the fully correlated \textit{ab initio} quantum dynamics of our system, we solve the many-body Schrödinger equation $(i \hbar \partial_t -\hat{\mathrm{H}})|\Psi(t)\rangle =0$ using the multilayer multiconfiguration time-dependent Hartree method for atomic mixtures (ML-MCTDHX) ~\cite{CaoBolsinger2017}. This {\it ab initio} method employs a time-dependent basis set of orthonormal states, truncating the many-body Hilbert space to the relevant part of the dynamics at each time step. The basis and the corresponding expansion coefficients are variationally optimized, which allows us to take intraspecies and interspecies correlations into account in a numerically efficient manner. 
In a first step, the many-body wave function $|\Psi(t)\rangle$ is expanded with respect to $\mathcal{M}$ orthonormal species functions $\{|\Phi_k^\sigma(t)\rangle\}_{k=1}^{\mathcal{M}} $ for each spin state $\sigma \in \{\uparrow,\downarrow\}$
\begin{equation}
   |\Psi(t)\rangle = \sum_{k=1}^\mathcal{M} \sqrt{\lambda_k(t)}|\Phi^\uparrow_k(t)\rangle |\Phi^\downarrow_k(t)\rangle \, .
   \label{Mlx:toplayer}
   \end{equation}
The expansion coefficients $\lambda_k(t)$ are the eigenvalues of the $\sigma$-component reduced density matrix. The latter is defined as $\rho_\sigma^{N_\sigma}(\vec{x},\vec{x}^\prime)=\langle \Psi(t) |\left[\prod_{i=1}^{N_\sigma}\psi_{\sigma}(x_i) \right]^\dagger \prod_{i=1}^{N_\sigma}\psi_{\sigma}(x^\prime_i)|\Psi(t)\rangle$.
In the second step, we expand the species functions with respect to a time-dependent number state basis $\{|\vec{n}^\sigma(t)\rangle\}$:
\begin{equation}
|\Phi_k^{\sigma}(t)\rangle = \sum_{\vec{n}} B^{\sigma}_{k; \vec{n}}(t) |\vec{n}^{\sigma}(t)\rangle \,,
\label{Mlx:secondlayer}
\end{equation}
with the corresponding expansion coefficients $B^{\sigma}_{k; \vec{n}}(t)$. The distinct number states $|\vec{n}^{\sigma}(t)\rangle$ refer to different occupation numbers $\vec{n}^{\sigma}=(n^{\sigma}_1,\ldots,n^{\sigma}_{m_\sigma})$ and read as
\begin{equation}
     |\vec{n}^{\sigma}(t)\rangle = \left[ \prod_{i=1}^{m_\sigma} \hat{a}^{n^{\sigma}_i}_{i, \sigma}(t)\right]^\dagger |0\rangle .
    \label{Mlx:numberstates}
\end{equation}
The creation operator $\hat{a}^{\dagger}_{i, \sigma}(t) $ creates a fermion in the time-dependent, variationally optimized single-particle function (SPF) $|\phi_i^\sigma (t) \rangle $:
\begin{equation}
    |\phi_{i}^\sigma (t) \rangle = \hat{a}^{\dagger}_{i,\sigma}(t)|0\rangle.
    \label{Mlx:spf}
\end{equation}
Those operators fulfill the fermionic anticommutation relations implying that ML-MCTDHX takes the appropriate particle exchange symmetry into account. Finally, the SPF are expanded with respect to a time-independent primitive basis set which corresponds to a $\mathcal{D}$-dimensional sine discrete variable presentation $\{|s\rangle\}$, where we use $\mathcal{D}=120$ grid points. Thus, this expansion reads as
\begin{equation}
    |\phi_{i}^\sigma(t) \rangle =\sum_{j=1}^{\mathcal{D}} b_{ij}^\sigma(t) |s_j\rangle.
    \label{Mlx:bottomlayer}
\end{equation}
Note that on this layer, the expansion coefficients are time-dependent while the basis set is time-independent.

By employing this truncation procedure, the determination of the total many-body wave function reduces to finding the expansion functions and respective coefficients of each layer at each individual time-step instead of using the full time-independent basis set, spanned by the primitive basis $\{ | s_j \rangle \}_{j = 1}^{\mathcal{D}}$. The time-evolution of the many-body wave equation is determined by solving the equations of motion of ML-MCTDHX, determined by the Dirac-Frenkel variational principle. The configuration space $(\mathcal{M},\,m_\sigma)_{\sigma \in\{\uparrow,\downarrow \}}$ defines the Hilbert space truncation ~\cite{CaoBolsinger2017}, here  $\mathcal{M}=m_\uparrow=m_\downarrow =6$.

\section{NONINTERACTING DESCRIPTION OF TRIPLE-WELL CONFINED FERMIONS}
\label{sec:non-interacting-effective}
We establish the specifics of the effective description of our triple-well system via an effective tight-binding model. To achieve this, we elaborate on the symmetry analysis of $\hat{H}_{\uparrow} + \hat{H}_{\downarrow}$ [Eq.~\eqref{Hamilton_mlx_bare}], outlined in Sec.~\ref{sec:effectivemodel_hamilton}.

\begin{figure}
    \centering
    \includegraphics[width=1.0\columnwidth]{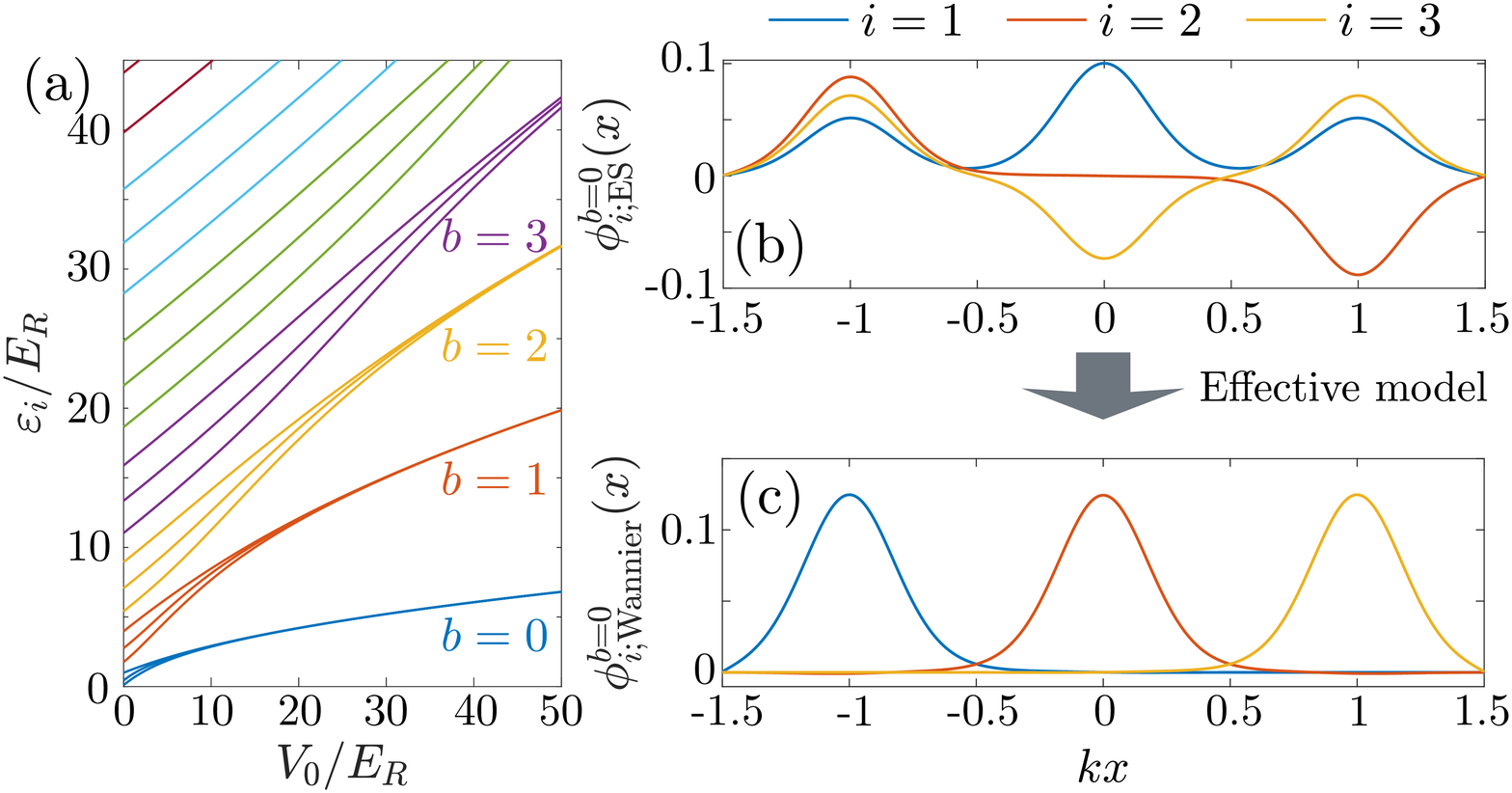}
    \caption{(a) The $21$ energetically lowest non-interacting eigenenergies of a triple-well with varying $V_0$. Eigenenergies belonging to different bands are indicated by lines of different color (see labels). (b) The $3$ energetically lowest eigenstates corresponding to the $b=0$ band of a triple-well with $V_0 = 12 E_{\rm R}$. (c) The $3$ Wannier states generated from the eigenstates of (b) via the procedure described in Appendix~\ref{sec:non-interacting-effective}.}
    \label{fig:non_interacting}
\end{figure}

Strictly speaking, no Bloch bands exist in our system due to the presence of hard-wall boundary conditions. However, we expect some precursors of the band-structure to persist especially in the case of large $V_0$.
Indeed, the explicit numerical diagonalization of Eq.~\eqref{Hamilton_mlx_bare} [see Fig.~\ref{fig:non_interacting}(a)] indicates that for deep lattices, $V_0 \geq 10 E_{\rm R}$, the low-lying states of the system organize in groups of three states analogously to our expectation for periodic boundary conditions.
We can further expect (and indeed we observe in Fig.~\ref{fig:non_interacting}(b)) that within each such group, herewith referred to as a band, the involved states are superpositions of states that are well-localized within the wells. The latter we call Wannier states, in analogy to the translationally invariant case.
To extract these states, see Fig.~\ref{fig:non_interacting}(c), we derive an effective model as follows. We first consider the most general single-particle tight-binding model with three sites that reads
\begin{equation}
\begin{split}
    \hat{H} = \sum_{b = 0}^{\infty} 
    \bigg(
        &- t_{LM}^b \hat{a}_{L}^{\dagger} \hat{a}_{M}
        - (t_{LM}^b)^* \hat{a}_{M}^{\dagger}\hat{a}_{L} \\
        &- t_{MR}^b \hat{a}_{M}^{\dagger} \hat{a}_{R}
        - (t_{MR}^b)^* \hat{a}_{R}^{\dagger}\hat{a}_{M} \\
        &+ \epsilon_{L}^b \hat{a}_{L}^{\dagger} \hat{a}_{L}
        + \epsilon_{M}^b \hat{a}_{M}^{\dagger} \hat{a}_{M} \\
        &+ \epsilon_{R}^b \hat{a}_{R}^{\dagger} \hat{a}_{R}
    \bigg),
\end{split}
\label{efectiveH_appendix}
\end{equation}
where $b$ refers to the index of the involved band. Here we have suppressed the spin index since it is not relevant in the discussion of this section. This Hamiltonian can be significantly simplified by using the symmetry properties of the total Hamiltonian, see Eq.~\eqref{Hamilton_mlx_bare}. Notice that Eq.~\eqref{Hamilton_mlx_bare} is real and thus the maximally localized Wannier states would also be real, thus,  $t_{LM}^b = t_({LM}^b)^*$ and $t_{MR}^b = (t_{MR}^b)^*$ holds. In addition, the potential is parity symmetric $x \to -x$ allowing to write $t_{LM}^b = t_{MR}^b = t^b$ and $\epsilon_{L}^b = \epsilon_{R}^b$ by demanding that Eq.~\eqref{efectiveH_appendix} also respects this symmetry. By defining $\epsilon_{M}^b = \epsilon^b$ and $\epsilon_{L}^b = \epsilon^b + \Delta \epsilon^b$, we arrive at 
\begin{equation}
\begin{split}
    \hat{H} = \sum_{b = 0}^{\infty} 
    \bigg[
        &- t^b \bigg( \hat{a}_{L}^{\dagger} \hat{a}_{M}
        + \hat{a}_{M}^{\dagger}\hat{a}_{R} + \text{H.c.} \bigg) \\
        &+ \left( \epsilon^{b} + \Delta \epsilon^b \right) 
        \left( \hat{a}_{L}^{\dagger} \hat{a}_{L}
        + \hat{a}_{R}^{\dagger} \hat{a}_{R} \right) \\
        &+ \epsilon^b \hat{a}_{M}^{\dagger} \hat{a}_{M}
    \bigg].
\end{split}
\end{equation}
The effective Hamiltonian depends on exactly three parameters for each band, $t^b$, $\epsilon^b$, and $\Delta \epsilon^b$, and within each band we have exactly three states. Thus, we can fit the effective model parameters so that it exactly reproduces the eigenenergies, $\varepsilon_i^b$, with $i = 1,2,3,$ of Eq.~\eqref{Hamilton_mlx_bare}, yielding
\begin{equation}
\begin{split}    
    \epsilon^b &= \varepsilon_1^b + \varepsilon_3^b - \varepsilon_2^b, \\
    \Delta \epsilon^b &= 2\varepsilon_2^b -(\varepsilon_1^b + \varepsilon_3^b), \\
    t^b &=\sqrt{\frac{1}{2}(\varepsilon_2^b -\varepsilon_1^b)(\varepsilon_3^b - \varepsilon_2^b)}.
\end{split}
\end{equation}
\begin{widetext}
With this choice, the localized basis is related to the energy eigenstates via the unitary transformation
\begin{equation}
    \hat{U}^b = \left(
    \begin{array}{c c c}
       \dfrac{2 t^b}{\sqrt{(\Delta \epsilon^b+ \Omega^b)^2 + 8 (t^b)^2}} & \dfrac{\Delta \epsilon^b+ \Omega^b}{\sqrt{(\Delta \epsilon^b+ \Omega^b)^2 + 8 (t^b)^2}} & \dfrac{2 t^b}{\sqrt{(\Delta \epsilon^b+ \Omega^b)^2 + 8 (t^b)^2}} \\
       \dfrac{1}{\sqrt{2}} & 0 & -\dfrac{1}{\sqrt{2}} \\
       \dfrac{2 t^b}{\sqrt{(\Delta \epsilon^b- \Omega^b)^2 + 8 (t^b)^2}} & \dfrac{\Delta \epsilon^b- \Omega^b}{\sqrt{(\Delta \epsilon^b- \Omega^b)^2 + 8 (t^b)^2}} & \dfrac{2 t^b}{\sqrt{(\Delta \epsilon^b- \Omega^b)^2 + 8 (t^b)^2}}
    \end{array}
    \right),
    \label{unitary_Wannier}
\end{equation}
where $\Omega^b \equiv \sqrt{(\Delta \epsilon^b)^2 + 8 (t^b)^2}$.
\end{widetext}
By inverting the transformation, we obtain the localized states in terms of the energy eigenstates which can be determined directly by diagonalizing Eq.~\eqref{Hamilton_mlx_bare}. The above procedure defines the Wannier states up to a constant phase shift. In order to remove this ambiguity, we fix the phase of the eigenstates of Eq.~\eqref{Hamilton_mlx_bare} by multiplying with the appropriate phase factor of $\pm 1$ so that they take positive values in the leftmost part of the triple-well $x \approx -3 \pi/2$. This is done because the left site is assumed to contribute with a positive phase to the energy eigenstates [see Eq.~\eqref{unitary_Wannier}].

\section{EFFECTIVE INTERACTIONS}
\label{app:effective}
In this appendix we outline the derivation of the interaction Hamiltonian [see Eq.~\eqref{effectiveH_interact_equ}] that is part of the effective model $\hat{H}_{tJU} =\hat{H}_0 +\hat{H}^\prime_I$. To this end, we consider the total many-body Hamiltonian from Eqs.~\eqref{Hamilton_mlx_bare} and \eqref{Hamilton_mlx_int}, $\hat{H}=\sum_{\sigma\in\{\uparrow,\downarrow\}} \hat{H}_\sigma + \hat{H}_I$. In our system, the gap between two energy bands constitutes the largest energy scale. Thus, it is justified that a corresponding tight-binding model might well capture the dynamics and for sufficiently small interactions, the many-body spectrum can be captured by the non-interacting single-particle eigenenergies (SPEE) of the potential. Accordingly, the many-body eigenstates can be described by the single-particle eigenstates (SPES). We transform these states to the Wannier states $\phi_s^b(x) $ describing the state on an energy band $b$ inside a well $s$ (see Appendix~\ref{sec:non-interacting-effective}). The underlying approximation is that the latter are well localized within the wells such that interactions of particles residing in distinct lattice sites can be neglected. 

Let us define the corresponding creation and annihilation operators of spin species $\sigma \in \{\uparrow, \downarrow\}$,
\begin{equation}
    \label{eq:creationann_operators_effective}
    \begin{split}
         &\hat{a}_{s \sigma}^{b\dagger} = \int \mathrm{d}x \phi_s^b(x) \hat{\psi}_{\sigma}^\dagger \\
        &\hat{a}_{s \sigma}^b = \int \mathrm{d}x \phi_s^{b*}(x) \hat{\psi}_{\sigma},
    \end{split}
\end{equation}
which create and annihilate a particle in the state $\phi_s^b(x) $ with eigenenergy $\epsilon_s^b$ and obey the fermionic anti-commutation relations. With these operators, the many-body Hamiltonian can be written as 
\begin{equation}
\label{eq:mb_Hamilton_rewritten}
    \begin{split}
     \hat{H}= \sum_{s\in\{\mathrm{L}, \mathrm{M},\mathrm{R}\}} \Bigl(\sum_b  \sum_{\sigma\in\{\uparrow,\downarrow\}} \epsilon_s^b \hat{a}_{s\sigma}^{b\dagger}\hat{a}_{ s\sigma}^b \\ + g  \sum_{b,b\prime,c,c^\prime} U_s^{bb^\prime c c^\prime} \hat{a}_{s\uparrow}^{b\dagger}\hat{a}_{s\downarrow}^{b^\prime\dagger}\hat{a}_{s\downarrow}^c\hat{a}_{s\uparrow}^{c^\prime} \Bigr),
    \end{split}
\end{equation}
with $ U_s^{bb^\prime c c^\prime} =\int \mathrm{d}x \phi_s^{c*}(x)\phi_s^{c^\prime*}(x)\phi_s^b(x)\phi_s^{b^\prime}(x)$. Let us analyze which interaction terms emanate from the last sum. First, we consider $b=b^\prime=c=c^\prime$ which yields an interaction term $\hat{H}^{doub}_s=g\sum_b U_s^{bbbb}\hat{a}_{s\uparrow}^{b\dagger}\hat{a}_{s\downarrow}^{b\dagger}\hat{a}_{s\downarrow}^b\hat{a}_{s\uparrow}^b $ and describes the doublon interaction energy when two particles of anti-oriented spins occupy the same state inside a well $s$. By considering the definition of $U$, we find that $ U_s^{bbbb} =\int \mathrm{d}x |\phi_s^b(x)|^4$. Subsequently, we abbreviate $U_s^{bbbb}$ as $U_s^{b}$. Next, we consider direct interactions between two particles of opposite spin in different states, which is defined by $\hat{H}^{dir}_s= g\sum_{b\neq b^\prime} U_s^{bb^\prime b^\prime b }\hat{a}_{s\uparrow}^{b\dagger}\hat{a}_{s\downarrow}^{b^\prime\dagger}\hat{a}_{s\downarrow}^{b^\prime}\hat{a}_{s\uparrow}^b$.
We rename the interaction constant as $J_s^{bb^\prime} = U_s^{bb^\prime b^\prime b }= \int \mathrm{d} x |\phi_s^b(x)|^2 |\phi_s^{b^\prime}(x)|^2$. With these two types of interactions, we can introduce a direct interaction Hamiltonian:
\begin{equation}
    \hat{H}_{I,dir} = g\sum_{s\in\{\mathrm{L}, \mathrm{M},\mathrm{R}\}} \Bigl( \sum_b U_s^b \hat{n}_{ s\uparrow}^b\hat{n}_{s \downarrow}^b+ \sum_{b\neq b^\prime} J_s^{bb^\prime} \hat{n}_{s \uparrow}^b \hat{n}_{s \downarrow}^{b^\prime}\Bigr).
\end{equation}
However, this is not sufficient to generate an effective description of interactions as it can be demonstrated that the last term breaks the SU(2)-symmetry of the total Hamiltonian. Thus, it would yield incorrect eigenstates. In order to avoid this artificial symmetry breaking, we include additional terms. From $ U_s^{bb^\prime c c^\prime}$, we find a spin exchange term of the form $ \hat{H}^{exc}_s= g\sum_{b\neq b^\prime} U_s^{bb^\prime b b^\prime} \hat{a}_{s\uparrow}^{b\dagger}\hat{a}_{s\downarrow}^{b^\prime\dagger}\hat{a}_{s\downarrow}^b\hat{a}_{s\uparrow}^{b^\prime}$, which allows two anti-aligned fermions occupying different states to exchange their spin. Ultimately, we find the effective interaction Hamiltonian:
\begin{equation}
    \begin{split}
        \hat{H}^\prime_I &=  \sum_{s\in\{\mathrm{L}, \mathrm{M},\mathrm{R}\}} \hat{H}^{doub}_s +\hat{H}^{dir}_s + \hat{H}^{exc}_s \\
        & = g \sum_{s\in\{\mathrm{L}, \mathrm{M},\mathrm{R}\}} \Bigl[\sum_b U_s^b \hat{n}_{s \uparrow}^b\hat{n}_{ s\downarrow}^b \\ &- \sum_{b\neq b^\prime} J_s^{bb^\prime}\bigl( \hat{\mathbf{S}}_s^b \cdot \hat{\mathbf{S}}_s^{b^\prime} - \frac{1}{4} \hat{n}_s^b \hat{n}_s^{b^\prime} \bigr)\Bigr],
    \end{split}
    \label{eq:effectiveH_final}
\end{equation}
with $ J_s^{bb^\prime}=U_s^{bb^\prime b^\prime b }= U_s^{bb^\prime b b^\prime}$. The operators $ \hat{\mathbf{S}}_s$ are the spin operators as introduced in Sec.~\ref{sec:effectivemodel_hamilton}. In conclusion, we have obtained Eq.~\eqref{effectiveH_interact_equ}. This Hamiltonian induces ferromagnetic order for $g>0$ and anti-ferromagnetic order for $g<0$. Note that additional terms can be derived from the second sum of Eq.~\eqref{eq:mb_Hamilton_rewritten}. These include more involved doublon tunneling terms, cradle modes and density-mediated interactions, see \cite{KoutentakisPhD} for more details. However, opposite to the interaction terms introduced above, applying these additional ones to a many-body state causes an additional energy shift in the case of vanishing interactions. Thus, they only contribute to avoided crossings when the two distinct corresponding configurations cross one another within $tJU$ (see Fig.~\ref{spectrum} for an example of this process) or generate small energy shifts to the $tJU$ eigenenergies due to level repulsion when $\partial E_{i}/ \partial g \approx \partial E_{j}/ \partial g$ for two $tJU$ eigenstates $i$, $j$ that couple by these interaction terms.

\section{EFFECT OF DOUBLONS IN THE MANY-BODY DYNAMICS}
\label{app:doublons}
Throughout the manuscript we have claimed that doublon formation is negligible in our system. The purpose of this section is to provide theoretical arguments on why this behavior occurs. Doublon formation in the ground band $b=0$ is energetically prohibitive for $U \gg t$, where $U$ and $t$ are the characteristic energy scales of intrasite interaction and tunneling respectively \cite{JoerdensStrohmaier2008}, due to the formation of an energy gap among the states with one and zero doublons. For the parameters of our setup $U/t = g \bar{U}^0/t^0 \approx 55.4 g$, where $\bar{U}^0 = \frac{1}{3}\sum_{s \in \{ {\rm L}, {\rm M}, {\rm R} \}} U^0_s$, and thus ground band doublons are expected to exist only for very small interactions $g \approx 0$. This is in agreement with our findings of Fig.~\ref{tripletsinglets}(c), demonstrating non-negligible doublon occupation only for $g < 0.1$.

Doublon formation in the excited band $b = 1$ requires the spin-$\downarrow$ particle to be in this band. This can be reached by interaction driven spin-exchange processes which do not change the spatial distribution of the particles~\cite{KollerWall2016, Koutentakis2019} and are not affected by the presence of a sizable band gap for the considered value of $V_0$ that prohibits inter-band population transfer. 
Such a process corresponds to the spin-exchange term $\propto J^{01}_s$ in Eq.~\eqref{effectiveH_interact_equ}. The importance of this term when compared to interactions in the excited band is $g$ independent, specifically $\bar{J}^{01}/\bar{U}^{1} \approx 0.67$, where $\bar{J}^{01} = \frac{1}{3}\sum_{s \in \{ {\rm L}, {\rm M}, {\rm R} \}} J^{01}_s$, while $g \bar{U}^1/t^1 \approx 2.17 g$. By using the above we can work out the timescale of the transport relative to the tunneling which corresponds to the inverse of the fraction of the two-energy scales, $g \bar{U}^1/t^1$. This implies that spin-exchange processes become important at $g \bar{U}^1/t^1 > 1$ yielding $g = g_0 \approx 0.67$. In that case the interaction strength is already large $g_0 \bar{U}^1/t^1 \approx 1.49$ and thus doublon formation is expected to be negligible. This argumentation is corroborated by our findings in Fig.~\ref{tripletsinglets}(c), where a small population of $P_D(t)$ is observed in the region $0.1 < g < 2$ associated to the formation of higher band doublons.

Finally, let us comment on the fact that $h_s(t)$ can be readily corrected for doublon occupation within state-of-the-art experiments. Note that a comparison of Figs.~\ref{holecomparison}(a)--\ref{holecomparison}(c) with Figs.~\ref{holecomparison}(d)--\ref{holecomparison}(f) already reveals that its effect on altering the value of $h_s(t)$ is negligible. However, a way to experimentally verify this claim is important. The correction mechanism we propose is based on quantum microscopy identifying the position of all atoms in the system in a spin-resolved manner and the fact that the doublon formation is associated to pronounced triple occupation of a single site. The triple occupation can be deduced experimentally by the reduction of number of atoms in the other wells even in the case that single and triple occupations of a site are not directly experimentally distinguishable.

In the case of $b = 0$ doublons the triple occupation signature occurs since these states occur for weak interaction where the correlations among the lower and higher band particles are negligible. This means that the hole would tunnel in and out of the higher band state at the doublon position on a time-scale proportional to $\hbar/t^1$. In contrast the doublon formation occurs on a much longer timescale $\hbar/t^0$. Therefore, the shift in $h_s(t)$ due to doublon formation can be removed by neglecting the cases where a triple occupation occurs since in these cases one of the observed holes corresponds to the ground state one created due to the doublon. The artificial increase of $h_s(t)$ occurring when the position of the hole coincides with the doublon can be removed by studying the motion of the holes in the cases that triple occupation of a site occurs and assuming that the $b = 1$ hole tunnels as a non interacting particle.

A similar procedure for $b = 1$ is significantly easier. This is because 
after a spin exchange occurs that transfers the spin-$\downarrow$ particle to the excited band, all atoms in the ground band are in the spin-$\uparrow$ state and occupy one site each. This means that if a doublon forms in the excited band, it can be always identified by the triple occupation of a site (two for the doublon and one for the spin-$\uparrow$ particle of the lower band) and thus this contribution can be safely removed.

\bibliography{fermions}
\end{document}